\documentclass[aps,prd,reprint,twocolumn,superscriptaddress,longbibliography,nofootinbib,floatfix,showpacs]{revtex4-2}

\usepackage{graphicx}
\usepackage{amsmath,amssymb,amsfonts}
\usepackage{hyperref}
\usepackage{slashed}
\usepackage{bm}
\usepackage[usenames]{color}
\usepackage[T1]{fontenc} \usepackage[latin1]{inputenc}

\usepackage{listings}
\usepackage[scaled=0.85]{inconsolata}  

\newcommand{\be}{\begin{equation}}
\newcommand{\ee}{\end{equation}}
\newcommand{\ud}{\mathrm{d}}

\newcommand{\LCm}{{\scriptscriptstyle -}} 
\newcommand{\LCp}{{\scriptscriptstyle +}}
\newcommand{\LCpm}{{\scriptscriptstyle\pm}}

\begin{document}

\title{Scattered wave functions and worldline instantons for particle production in curved spacetime}

\author{Philip Semr\'en}
\email{philip.semren@umu.se}

\author{Greger Torgrimsson}
\email{greger.torgrimsson@umu.se}
\affiliation{Department of Physics, Ume{\aa} University, SE-901 87 Ume{\aa}, Sweden}

\begin{abstract}
We study the production of spin-$1/2$ particle-antiparticle pairs in curved spacetimes with nontrivial dependence on more than one coordinate. To this end, we develop two complementary approaches. First, we extend the scattered-wave-function (SWF) method, originally introduced for pair production in electromagnetic backgrounds, to curved spacetime backgrounds. Second, we complete the development of an open-worldline-instanton method by deriving the pre-exponential factor of the pair-production probability. We apply both methods to several two-dimensional metrics and find good agreement between the resulting probabilities. While the SWF approach provides numerically exact results and is particularly efficient for the examples considered here, the instanton approach offers favorable scaling to higher-dimensional backgrounds and more extreme parameter regimes. These methods provide new tools for studying pair production in multidimensional gravitational backgrounds beyond the reach of many existing approaches.
 
\end{abstract}

\maketitle

\section{Introduction}

Pair production in curved spacetimes~\cite{Parker:2009uva,Ford:2021syk} was originally studied in detail by Parker in the 60's~\cite{Parker:2025jef,Parker:1969au,Parker:1971pt} for time-dependent cosmological spacetimes, and by Hawking in the 70's for black holes~\cite{Hawking:1975vcx}.
Existing methods for pair production in curved spacetimes are still typically restricted to metrics that depend only on a single spacetime coordinate, e.g. a radial coordinate for black holes or time for Friedmann-Lema\^itre-Robertson-Walker metrics. In this paper we will present two new approaches that can both deal with spacetimes that depend on more than one coordinate. 

The first approach is exact at leading order in the Furry-picture expansion and is based on scattered wave functions (SWF). It was originally developed for QED processes in strong electromagnetic (EM) background fields $F_{\mu\nu}$~\cite{Torgrimsson:2025pao,Torgrimsson:2025ihd} and has proven very efficient when implemented on a GPU, enabling the first studies of backgrounds depending on three or all four coordinates, $F_{\mu\nu}(t,x,y)$ or $F_{\mu\nu}(t,x,y,z)$. 

For comparison, two of the historically most widely used numerical methods for Schwinger pair production, the Dirac-Heisenberg-Wigner method~\cite{Bialynicki-Birula:1991jwl} and the CQFT method~\cite{Jiang:2012mfn}, have recently been extended from EM to curved backgrounds with dependence on a single spacetime coordinate~\cite{Semren:2024fkb,Alkhateeb:2026heb}. In the QED case, several numerical approaches have been applied to $F_{\mu\nu}(t,x)$ backgrounds in a large number of studies over the past decade; see e.g.~\cite{Hebenstreit:2011wk,Jiang:2012mfn,Jiang:2013wha,Wollert:2015kra,Kohlfurst:2015zxi,Kohlfurst:2015niu,Aleksandrov:2016lxd,Aleksandrov:2017mtq,Kohlfurst:2017hbd,Kohlfurst:2017git,Lv:2018wpn,Peng:2018hmj,Ababekri:2019dkl,Kohlfurst:2019mag,Aleksandrov:2019ddt,Ababekri:2019qiw,Li:2021vjf,Mohamedsedik:2021pzb,Li:2021wag,Kohlfurst:2021skr,Jiang:2023hbo}. It seems likely that at least some of these methods can be extended to curved spacetimes that depend on two coordinates, $g_{\mu\nu}(t,x)$. Indeed, much of the framework required for such an extension within the CQFT approach appears to have already been developed in~\cite{Alkhateeb:2026heb}. However, despite the obvious motivation to go beyond $F_{\mu\nu}(t,x)$ backgrounds, they have remained the state of the art for more than a decade.

The second approach provides a semiclassical approximation and is based on {\it open} worldline instantons\footnote{Open worldlines were used to study pair production by a constant electric field in~\cite{Barut:1989mc,Rajeev:2021zae}.}. Like the SWF approach, it was originally developed for pair production in EM fields~\cite{DegliEsposti:2021its,DegliEsposti:2022yqw,DegliEsposti:2023qqu,DegliEsposti:2023fbv,DegliEsposti:2024rjw,DegliEsposti:2024upq}, and has also been applied to $F_{\mu\nu}(t,x,y,z)$ fields~\cite{DegliEsposti:2023qqu}. There exists an older, {\it closed} worldline-instanton approach that was developed already in~\cite{Affleck:1981bma,Dunne:2005sx,Dunne:2006st,Dunne:2006ur}. It is a powerful method for obtaining integrated probabilities, but it does not give the momentum or spin dependence. 

Moreover, the numerical methods in~\cite{Hebenstreit:2011wk,Jiang:2012mfn,Jiang:2013wha,Wollert:2015kra,Kohlfurst:2015zxi,Kohlfurst:2015niu,Aleksandrov:2016lxd,Aleksandrov:2017mtq,Kohlfurst:2017hbd,Kohlfurst:2017git,Lv:2018wpn,Peng:2018hmj,Ababekri:2019dkl,Kohlfurst:2019mag,Aleksandrov:2019ddt,Ababekri:2019qiw,Li:2021vjf,Mohamedsedik:2021pzb,Li:2021wag,Kohlfurst:2021skr,Jiang:2023hbo} are constructed in a way that naturally yields probability spectra as functions of a single momentum variable, i.e. $P({\bf p})$, $P({\bf q})$ or $P({\bf p}-{\bf q})$, where ${\bf p}$ and ${\bf q}$ are the momenta of the particle and antiparticle. By contrast, both approaches presented here yield the joint probability $P({\bf p},{\bf q})$, thereby retaining the full particle--antiparticle momentum correlation. 

We recently initiated the extension of the open-worldline-instanton approach from QED to curved spacetime in~\cite{Semren:2025dix}, where the exponential contribution to the pair-production probability was obtained\footnote{Worldlines were also recently used in~\cite{Ilderton:2025umd} to obtain wave functions in a Vaidya metric.}. In the present work we derive the corresponding pre-exponential factor, thereby completing the leading-order semiclassical formulation. To do this, we first derive an exact worldline path-integral representation of the propagator in general curved spacetimes and EM fields, which then serves as the starting point for deriving the worldline-instanton formulas. 

The standard tunneling approach to pair production by black holes~\cite{Srinivasan:1998ty,Shankaranarayanan:2000qv,Parikh:1999mf,Vanzo:2011wq} is applicable to cases where a certain 1D integral can be performed with the residue method, with a residue at the horizon that gives an imaginary part in the exponent and hence Hawking radiation. For the multidimensional backgrounds considered here, the instanton calculation cannot be reduced to such a 1D residue integral. 

Throughout this paper we focus on metrics with nontrivial dependence on more than one coordinate. We refer to metrics of the form $g_{\mu\nu}(t,x)$, $g_{\mu\nu}(t,x,y)$ and $g_{\mu\nu}(t,x,y,z)$ as 2D, 3D, and 4D backgrounds, respectively, even though the spacetime indices continue to run over $\mu=0,1,2,3$ in each case. In this classification, static and spherically symmetric black holes $g_{\mu\nu}(r)$ and Friedmann-Lema\^itre-Robertson-Walker metrics $g_{\mu\nu}(t)$ are therefore 1D. 
We will apply both the SWF and the instanton approaches to a set of 2D metrics and compare the resulting pair-production probabilities, finding good agreement. 

We use units with $c=\hbar=m=1$ and absorb a factor of the charge into the definition of the EM background field, $eA_\mu\to A_\mu$. The EM field tensor is given by $F_{\mu\nu}=\partial_\mu A_\nu-\partial_\nu A_\mu$.
We use the same conventions as in~\cite{FosterNightingale} for the metric objects, i.e. the Minkowski metric is given by
$\eta_{ab}=\text{diag.}(1,-1,-1,-1)$, the Christoffel symbols by
\be
\Gamma^\mu_{\rho\sigma}=\frac{1}{2}g^{\mu\nu}(g_{\nu\rho,\sigma}+g_{\nu\sigma,\rho}-g_{\rho\sigma,\nu}) \;,
\ee
the Riemann tensor by
\be
R^\mu_{\alpha\beta\gamma}=\Gamma^\mu_{\alpha\gamma,\beta}-\Gamma^\mu_{\alpha\beta,\gamma}
+\Gamma^\mu_{\beta\rho}\Gamma^\rho_{\alpha\gamma}-\Gamma^\mu_{\gamma\rho}\Gamma^\rho_{\alpha\beta} \;,
\ee
$R_{\mu\alpha\beta\gamma}=g_{\mu\nu}R^\nu_{\alpha\beta\gamma}$, and the Ricci tensor and scalar by
\be
R_{\mu\nu}=R^\rho_{\mu\nu\rho}
\qquad
R=g^{\mu\nu}R_{\mu\nu} \;.
\ee
The metric determinant is
\be
g=-\text{det }g_{\mu\nu}>0 \;.
\ee
We assume spacetimes that are asymptotically flat, but which can have a nontrivial holonomy.

\section{Dirac equation in curved spacetime}

We treat the Dirac equation as described e.g. in~\cite{Parker:2009uva}, except we use a different sign convention. 
The coupling between spinors and the curved background is expressed in terms of a tetrad/vierbein $e^a_\mu(x)$,
\be
g_{\mu\nu}(x)=e^a_\mu(x)e^b_\nu(x)\eta_{ab} \;,
\ee
and its inverse $e_a^\mu$,
\be
e^a_\mu e_a^\nu=\delta_\mu^\nu
\qquad
e^a_\mu e_b^\mu=\delta^a_b \;.
\ee
We use Greek letters for spacetime indices and Latin letters from the beginning of the alphabet for local Lorentz indices. Specific local Lorentz components are denoted by hatted numbers:
\be
\begin{aligned}
\mu&=0,1,2,3 &\qquad a&=\hat{0},\hat{1},\hat{2},\hat{3} \\
i&=1,2,3 &\qquad \hat{i}&=\hat{1},\hat{2},\hat{3} \;.
\end{aligned}
\ee
We have two different sets of gamma matrices: $\gamma^a$ are constant, while $\gamma^\mu$ depend on $x^\mu$ as
\be
\{\gamma^a,\gamma^b\}=2\eta^{ab}
\quad
\gamma^\mu=e^\mu_a\gamma^a
\quad
\{\gamma^\mu,\gamma^\nu\}=2g^{\mu\nu} \;.
\ee
The Dirac equation is given by
\be\label{Dirac1}
(i\gamma^\mu D_\mu-1)\psi=0 \;,
\ee
where the background covariant derivative,
\be\label{covariantDer}
D_\mu=\partial_\mu-\frac{i}{4}\omega_{\mu ab}\sigma^{ab}+iA_\mu
\qquad
\sigma^{ab}=\frac{i}{2}[\gamma^a,\gamma^b] \;,
\ee
contains both the EM gauge field $A_\mu$ and the spin connection (Ricci rotation coefficients)
\be\label{connection}
\begin{split}
\omega_{\mu ab}&=-\omega_{\mu ba}=e_{a\lambda}\left[\partial_\mu e^\lambda_b+\Gamma^\lambda_{\mu\nu}e^\nu_b\right] \\
\omega_{abc}&=e_a^\mu\omega_{\mu bc} \;.
\end{split}
\ee

By multiplying~\eqref{Dirac1} with $i\gamma^0$ and using $(\gamma^0)^2=g^{00}$, we can rewrite the Dirac equation in a form suitable for numerical integration,
\be\label{dt_Dirac}
\partial_t\psi=-\left[iA_0+\alpha^k(\partial_k+iA_k)+W\right]\psi \;,
\ee
where
\be
\alpha^k=\frac{1}{g^{00}}\gamma^0\gamma^k
\ee
and
\be
W=\frac{1}{4g^{00}}\omega_{abc}\gamma^0\gamma^a\gamma^b\gamma^c+\frac{i}{g^{00}}\gamma^0 \;.
\ee

The spin-connection term can be decomposed into a vector and a pseudo-vector as (see e.g.~\cite{Aldrovandi:2013wha})
\be
\frac{1}{4}\omega_{abc}\gamma^a\gamma^b\gamma^c=V_a\gamma^a+\tilde{V}_a\gamma^a\gamma^{\hat{5}} \;,
\ee
where $\gamma^{\hat{5}}=i\gamma^{\hat{0}}\gamma^{\hat{1}}\gamma^{\hat{2}}\gamma^{\hat{3}}$ and
\be
V_a=\frac{1}{2}{\omega^b}_{ba}
\qquad
\tilde{V}^a=-\frac{i}{4}\epsilon^{abcd}\omega_{bcd} \;.
\ee
This allows one to rewrite~\eqref{dt_Dirac} as
\be\label{dt_Dirac_V}
\begin{split}
\partial_t\psi=-\Big[&iA_0+V_0+\tilde{V}_0\gamma^{\hat{5}}\\
+\alpha^k(\partial_k+&iA_k+V_k+\tilde{V}_k\gamma^{\hat{5}})+\frac{i}{g^{00}}\gamma^0\Big]\psi \;.
\end{split}
\ee
This form suggests some degree of symmetry between the coupling to the EM field and the metric, especially for 2D metrics, $g_{\mu\nu}(t,x)$ with $\mu=0,1,2,3$, for which the pseudo-vector part vanishes, $\tilde{V}_\mu=0$. However, the metric also enters in the matrices $\alpha^k$ and $\gamma^0/g^{00}$. 

In practice, we can write a Mathematica or SymPy code that takes as input an analytical expression for $e^a_\mu$ and outputs simplified analytical expressions for all the terms on the right-hand side of~\eqref{dt_Dirac} or~\eqref{dt_Dirac_V}, and then one will obtain the same simplified expressions regardless of whether one starts with~\eqref{dt_Dirac} or~\eqref{dt_Dirac_V}. 

The observables are invariant under three different types of transformations:
\begin{itemize}
\item EM gauge transformation
\be\label{gauge1}
A_{\rm new}^\mu=A_{\rm old}^\mu+\partial^\mu f    
\ee
\item change of coordinates (diffeomorphism)
\be\label{gauge2}
\begin{split}
x_{\rm old}^\mu&=f^\mu(x_{\rm new}) \\
{{e_{\rm new}}^a}_\mu(x_{\rm new})&={{e_{\rm old}}^a}_\nu(f[x_{\rm new}]){f^\nu}_{,\mu}    
\end{split}
\ee
\item local Lorentz transformation
\be\label{gauge3}
{{e_{\rm new }}^a}_\mu(x)={\Lambda^a}_b(x){{e_{\rm old }}^b}_\mu(x) \;,
\ee
where
\be
\Lambda^T\eta\Lambda=\eta \;.
\ee
\end{itemize}

Because of~\eqref{gauge3}, $e^a_\mu$ is not uniquely determined by $g_{\mu\nu}$, but for a 2D metric of the form
\be
\ud s^2=a(t,x)\ud t^2+2c(t,x)\ud t\ud x-b(t,x)\ud x^2-\ud y^2-\ud z^2
\ee
one can choose
\be\label{e2Dtriang}
{e^a}_\mu=\begin{pmatrix}\frac{\sqrt{ab+c^2}}{\sqrt{b}}&0&0&0\\
-\frac{c}{\sqrt{b}}&\sqrt{b}&0&0\\
0&0&1&0\\
0&0&0&1
\end{pmatrix} \;.
\ee

We can check that the methods presented below have been correctly implemented in the code by checking that the resulting probability spectrum remains invariant under the transformations in~\eqref{gauge1}, \eqref{gauge2} and~\eqref{gauge3}.


\subsection{Tetrad gauge choice}

To construct a tetrad automatically from a metric, we must choose a gauge. A common choice is Schwinger's time gauge~\cite{Schwinger:1963re}
\be\label{timeGauge}
e^{\hat{0}}_j=0 \;,
\ee
which implies $e_{\hat{k}}^0=0$, $e_{\hat{0}}^0=1/e_0^{\hat{0}}$
and
\be
e^{\hat{k}}_j e^j_{\hat{l}}=\delta^k_l
\qquad
e^i_{\hat{k}} e^{\hat{k}}_j=\delta^i_j \;.
\ee
The spatial metric can then be expressed entirely in terms of the spatial components of the tetrad, $-g_{ij}=e^{\hat{k}}_i e^{\hat{l}}_j\delta_{ij}$
or
\be
-{\bf g}={\bf e}^T{\bf e} \;,
\ee
where ${\bf e}_{kj}=e^{\hat{k}}_j$. However, the condition in~\eqref{timeGauge} fixes only three of the six gauge degrees of freedom associated with the local Lorentz transformation. The remaining three can be used to make ${e^a}_\mu$ lower triangular. 

We assume that the hypersurfaces $x^0=\text{const.}$ are spacelike.
In the $(+---)$ convention this is equivalent to the spatial metric
$-{\bf g}$ being positive definite. The Cholesky decomposition then provides a unique lower-triangular
matrix ${\bf e}$ with positive diagonal entries satisfying $-{\bf g}={\bf e}^T{\bf e}$. Together with~\eqref{timeGauge}, this leads to a fully lower-triangular tetrad gauge (see e.g.~\cite{Mei:2007xk,Thiemann:2023lcy}),
\be
{e^a}_\mu=\begin{pmatrix}
\cdot&0&0&0\\
\cdot&\cdot&0&0\\
\cdot&\cdot&\cdot&0\\
\cdot&\cdot&\cdot&\cdot
\end{pmatrix} \;.
\ee
The nonzero components can be expressed in terms of the minors of $g_{\mu\nu}$, defined by
\be
[g]_{M,N}=\det{g_{M,N}} \;,
\ee
where $g_{M,N}$ denotes the submatrix of $g_{\mu\nu}$ obtained by retaining the rows indexed by $M$ and the columns indexed by $N$, so that e.g.
\be
[g]_{123,023}=\det\begin{pmatrix}
 g_{10} & g_{12} & g_{13} \\
 g_{20} & g_{22} & g_{23} \\
 g_{30} & g_{32} & g_{33}
\end{pmatrix} \;.
\ee
Then (cf.~\cite{Mei:2007xk})
\be
\begin{split}
{e^{\hat{0}}}_\mu&=\left(\frac{\sqrt{-[g]_{0123,0123}}}{\sqrt{-[g]_{123,123}}},0,0,0\right)\\
{e^{\hat{1}}}_\mu&=-\frac{([g]_{123,023},[g]_{123,123},0,0)}{\sqrt{-[g]_{123,123}}\sqrt{[g]_{23,23}}}\\
{e^{\hat{2}}}_\mu&=\frac{([g]_{23,03},[g]_{23,13},[g]_{23,23},0)}{\sqrt{[g]_{23,23}}\sqrt{-g_{33}}}\\
{e^{\hat{3}}}_\mu&=-\frac{(g_{30},g_{31},g_{32},g_{33})}{\sqrt{-g_{33}}} \;.
\end{split}
\ee
For 2D metrics, this reduces to~\eqref{e2Dtriang}.

We already know that $-[g]_{0123,0123}=-\det g_{\mu\nu}>0$. The reality of the other square roots follows from Sylvester's criterion for
positive-definite matrices applied to $-g_{ij}$. For this particular case, the criterion can be proved as follows. Since $-{\bf g}$ is positive definite,
\be
(u,v,w)\cdot(-{\bf g})\cdot(u,v,w)>0
\ee
for all nonzero $(u,v,w)$. Cancelling the mixed terms by successive shifts,
\be
w\to w-\frac{g_{31}u+g_{32}v}{g_{33}} 
\qquad
v\to v+\frac{g_{31}g_{32}-g_{21}g_{33}}{g_{22}g_{33}-g_{32}^2}u \;,
\ee
gives
\be
-g_{33}w^2
+\frac{g_{22}g_{33}-g_{32}^2}{-g_{33}}v^2
+\frac{-\det g_{ij}}{g_{22}g_{33}-g_{32}^2}u^2>0 \;.
\ee
Since the inequality holds for arbitrary $u,v,w$, each coefficient
must be positive, implying
\be
-g_{33}>0,\qquad
g_{22}g_{33}-g_{32}^2>0,\qquad
-\det g_{ij}>0 \;.
\ee
Thus, all square roots appearing in the tetrad are real whenever the
$x^0=\mathrm{const.}$ hypersurfaces are spacelike.

\subsection{Holonomy}

The methods we present here can be used for 3D and 4D metrics, but we will focus on 2D examples below. While this makes the computations faster, 2D fields have a more nontrivial topology that sometimes prevents one from making simplifications that one would be able to do in 3D or 4D. The relevant topology here is that of the exterior (approximately flat-space) region, i.e. the region surrounding a compact interior region where spacetime is curved, 
\be
\begin{split}
\text{interior:}&\qquad |x^\mu|<L \qquad R^\mu_{\alpha\beta\gamma}\ne0\\
\text{exterior:}&\qquad |x^\mu|>L \qquad R^\mu_{\alpha\beta\gamma}\approx0 \;.
\end{split}
\ee
In 2D, the exterior region $|x^\mu|=\sqrt{t^2+x^2}>L$ is not simply connected, whereas in 3D or 4D it is -- so in 2D there are loops in the exterior that cannot be contracted to a point without leaving the exterior, in contrast to exterior loops in 3D and 4D. This topological distinction can be made precise in terms of the holonomy. 

We first note that the Riemann tensor can be expressed entirely in terms of the spin connection in the same way as ${\bf F}_{\mu\nu}$ is obtained from ${\bf A}_\mu$ for non-Abelian Yang-Mills fields,
\be
{\bf R}_{\mu\nu}=\partial_\mu{\bm\omega}_\nu-\partial_\nu{\bm\omega}_\mu+[{\bm\omega}_\mu,{\bm\omega}_\nu] \;,
\ee
where ${({\bm\omega}_\mu{\bm\omega}_\nu)^a}_b={{\omega_\mu}^a}_c{{\omega_\nu}^c}_b$ and ${{{\bf R}_{\mu\nu}}^a}_b=e^a_\rho e^\sigma_b R^\rho_{\sigma\mu\nu}$. Consider the Wilson line
\be
{\bm\Lambda}_0[x(s)]=\mathcal{P}\exp\left(-\int_0^1\ud s\,\dot{x}^\mu{\bm\omega}_\mu\right) \;,
\ee
where $x^\mu(s)$ is some path and $\mathcal{P}$ means path ordering. For a small closed loop,  $x^\mu(0)=x^\mu(1)$ and $|x^\mu(s)-x^\mu(0)|\ll1$, 
\be\label{LambdaR}
{\bm\Lambda}_0\approx1+\frac{1}{2}{\bf R}_{\mu\nu}[x^\mu(0)]\oint\ud x^\mu x^\nu \;.
\ee
This can be rewritten as a surface integral and the generalization to a large loop is given by the non-Abelian Stokes theorem~\cite{Arefeva}.  
It follows from~\eqref{LambdaR} that ${\bm\Lambda}_0$ is path independent within any simply connected subset of the exterior (where ${\bf R}_{\mu\nu}=0$), so
\be
{\bm\Lambda}_0(x^\mu)=\mathcal{P}\exp\left(-\int_x^{x_r}\!\ud x^\mu{\bm\omega}_\mu\right)
\ee
becomes a proper function of the starting point $x^\mu$ (with $x_r^\mu$ a fixed reference point). One can then differentiate to find
\be
{\bm\omega}_\mu(x)={\bm\Lambda}_0^{-1}\partial_\mu{\bm\Lambda}_0 \;.
\ee
Under the local Lorentz transformation~\eqref{gauge3}, or ${\bf e}_\mu^{\rm new}={\bm\Lambda}{\bf e}_\mu$, ${\bm\omega}_\mu$ transforms as a non-Abelian gauge field,
\be
{\bm\omega}_\mu^{\rm new}={\bm\Lambda}{\bm\omega}_\mu{\bm\Lambda}^{-1}+{\bm\Lambda}\partial_\mu{\bm\Lambda}^{-1} \;.
\ee
With ${\bm\Lambda}={\bm\Lambda}_0$ we find ${\bm\omega}_\mu^{\rm new}=0$. Thus, $R^\rho_{\sigma\mu\nu}=0$ implies that ${\bm\omega}_\mu$ is pure gauge, so that one can locally choose ${\bm\omega}_\mu=0$. Things are more complicated when the exterior is not simply connected. For simplicity, we will assume that one can choose ${\bm\omega}_\mu=0$ everywhere in the exterior without making $e^a_\mu$ multivalued. This is the case for the examples we have considered and it still allows us to consider metrics with nontrivial holonomy.    

The vectors $e_a=e_a^\mu\partial_\mu$ satisfy $[e_a,e_b]=f^c_{ab}e_c$
with anholonomy/structure coefficients given by\footnote{Compared to the notation in~\cite{Aldrovandi:2013wha}, we have $\omega_{abc}=A_{bca}$ and $f_{abc}^{\rm here}=f_{cab}^{\rm there}$.}
\be\label{structure}
f_{abc}=f^d_{ab}\eta_{dc}=-e^\mu_a e^\nu_b(\partial_\mu e_{\nu c}-\partial_\nu e_{\mu c}) \;.
\ee
By rewriting~\eqref{connection} as
\be\label{connection2}
\partial_\mu e_{\nu a}=\Gamma^\rho_{\mu\nu}e_{\rho a}+\omega_{\mu ba}e^b_\nu 
\ee
one finds the structure coefficients in terms of the spin connection or vice versa,
\be
\begin{split}
f_{abc}&=-\omega_{abc}+\omega_{bac} \\
\omega_{abc}&=\frac{1}{2}(-f_{abc}+f_{bca}-f_{cab}) \;.
\end{split}
\ee
In the exterior region we therefore have
\be\label{integrability}
\omega_{\mu ab}=0 \iff \partial_\mu e_\nu^a-\partial_\nu e_\mu^a=0 \;.
\ee
Poincare's lemma~\cite{Nakahara:2003nw} says that $\partial_\mu e_\nu^a=\partial_\nu e_\mu^a$ is equivalent to the existence of a {\it local} coordinate system $y^a(x)$ such that $\partial y^a/\partial x^\mu=e^a_\mu$ and $g_{\mu\nu}=\eta_{\mu\nu}$. 

Such coordinates can be constructed as
\be\label{yaInt}
y^a=\int_{x_r}^x\ud\tilde{x}^\mu e_\mu^a(\tilde{x})=\int_0^1\ud s\,\dot{x}^\mu(s)e_\mu^a[x(s)] \;,
\ee
where $x^\mu(s)$ is some path, in the exterior, from some fixed reference point $x_r^\mu$ to $x^\mu$. The integral is locally path independent due to Stokes theorem: The difference between integrating along two different paths, $\gamma_1$ and $\gamma_2$, is a loop integral over the $1$-form $e^a=e^a_\mu\ud x^\mu$, which can be converted to a surface integral as
\be
\Delta y^a=\oint e^a=\int_S\ud e^a \;,
\ee
where $S$ is some surface with the loop $\gamma_1-\gamma_2$ as boundary. The exterior derivative is obtained from the structure constants in~\eqref{structure} according to Cartan's structure equation
\be
\ud e^a=-\frac{1}{2}f^a_{bc}e^b\wedge e^c \;,
\ee
so if $\omega_{\mu ab}=0$ on $S$ then $\Delta y^a=0$ and the two paths give the same result. In 3D and 4D, we can always choose a surface $S$ that stays in the exterior where $\omega_{\mu ab}=0$, so that~\eqref{yaInt} provides coordinates that cover the entire exterior and that give $g_{ab}=\eta_{ab}$ there. In contrast, in 2D, the exterior region has a hole, so two paths that go around the hole clockwise and anticlockwise can give different results. In other words, the holonomy of the tetrad $e_\mu^a$ is nontrivial. This is the basic fact that a closed form ($\ud e^a=0$) is only guaranteed to be exact ($e^a=\ud y^a$) locally, by Poincar\'e's lemma, while globally there can be an obstruction due to a nontrivial holonomy/De Rham cohomology~\cite{Nakahara:2003nw}. If $\Delta y^a\ne0$ then the coordinates constructed as~\eqref{yaInt} will be multivalued.

Thus, in 2D we may be interested in metrics which are asymptotically flat in all directions, but for which $g_{\mu\nu}\ne\eta_{\mu\nu}$ in the exterior. Such cases require additional care compared with the conceptually simpler situations encountered in 3D and 4D. Of course, not all 2D metrics have a nontrivial holonomy. Moreover, suppose we have a spacetime with a nontrivial holonomy for which we can choose coordinates so that $g_{\mu\nu}(t=\infty,x)=\eta_{\mu\nu}$ but then $g_{\mu\nu}(t=-\infty,x)=g_{\mu\nu}^{\rm past}(x)$ cannot be chosen as $\eta_{\mu\nu}$. Then it may still be possible to recover the same pair-production probability as an adiabatic limit of a spacetime with trivial holonomy. One way to construct such a limit is to introduce a sufficiently slow ramp in the far past, interpolating from $\lim_{t\to-\infty}g_{\mu\nu}=\eta_{\mu\nu}$, to $g_{\mu\nu}(t_p,x)=g_{\mu\nu}^{\rm past}(x)$ at some large negative time $t_p$, while leaving the metric unchanged for $t>t_p$. Essentially no pairs will be produced during the adiabatic ramp if it is slow enough. However, a slow ramp may not be numerically convenient.

\subsection{2D example 1}

As a first example, we consider metrics of the form (omitting the trivial coordinates $y$ and $z$)
\be\label{inflation2D}
\ud s^2=\ud t^2-e^{2A(t,x)}\ud x^2 \;.
\ee
We consider this metric primarily because it provides a convenient test case for our methods, but it can also be viewed as a toy model of a cosmological spacetime with localized inflation (or deflation). Pair production by a constant electric field in de Sitter space has been studied extensively; see e.g.~\cite{Stahl:2015gaa} for production of spin-$1/2$ particles. The methods developed here could equally well be applied to localized versions of such backgrounds, but for simplicity we set $A_\mu=0$ here.

The corresponding Ricci scalar is
\be\label{Rinflation2D}
R(t,x)=2\frac{\partial_t^2 e_{\hat{1}1}}{e_{\hat{1}1}}=2(A_t^2+A_{tt}) \;,
\ee
where $A_t=\partial_t A$.
In the gauge~\eqref{e2Dtriang}, the tetrad and spin connection are
\be
{e^a}_\mu=\begin{pmatrix}1&0\\0& e^A\end{pmatrix}
\quad
\omega_{\hat{1}ab}=\begin{pmatrix}0&A_t\\-A_t&0\end{pmatrix} 
\quad
\omega_{\hat{0}ab}=0 \;,
\ee
so the Dirac equation becomes
\be
\left(\partial_t+\frac{1}{2}A_t+\alpha^{\hat{1}}e^{-A}\partial_x+i\beta\right)\psi=0 \;,
\ee
where
\be
\beta=\gamma^{\hat{0}}
\qquad
\alpha^{\hat{i}}=\gamma^{\hat{0}}\gamma^{\hat{i}} \;.
\ee
In a 4D Friedmann-Lema\^itre-Robertson-Walker metric with the same scale factor multiplying all three spatial directions, we would have $\partial_t A(t,x)\to H(t)=\dot{a}/a$ and $1/2\to3/2$, as expected~\cite{Parker:1969au,Parker:1971pt}.

As a concrete example, consider
\be\label{Aex2D}
A(t,x)=\alpha\tanh(\omega t)\text{sech}^2(kx) \;.
\ee
Asymptotically, $\omega_{\mu ab}\to0$ and $\partial_\mu e_\nu^a=\partial_\nu e_\mu^a$, so we could locally define new coordinates in the exterior by~\eqref{yaInt} with e.g. $x_r^1=0$ and $t_r=10/\omega$ (so that $\tanh(\omega t_r)\approx1$). However, if we consider a loop around the interior, we find a nonzero holonomy
\be
\Delta y^{\hat{1}}=\oint\ud x^\mu e_\mu^{\hat{1}}=\int_{-\infty}^\infty\ud x\left[e^{A(\infty,x)}-e^{A(-\infty,x)}\right]\ne0 \;,
\ee
so that e.g. the value of $y^{\hat{1}}$ at the reference point is given by $y^{\hat{1}}(x_r)=n\Delta y^{\hat{1}}$ where $n$ is an integer. Thus, $y^{\hat{1}}(t,x)$ is a multivalued function.

\subsection{2D example 2}

As a second example, we consider metrics of the Gullstrand-Painlev\'e form,
\be\label{dsGP}
\ud s^2=\ud t^2-[\ud x-v(t,x)\ud t]^2 \;,
\ee
for which 
\be\label{Ricci_GP}
R=2(v_x^2+vv_{xx}+v_{tx})
\ee
and
\be
{e^a}_\mu=\begin{pmatrix} 1&0\\-v&1\end{pmatrix}
\quad
\omega_{\hat{1}ab}=\begin{pmatrix}0&v_x\\-v_x&0\end{pmatrix}
\quad
\omega_{\hat{0}ab}=0 \;,
\ee
so
\be
\left(\partial_t+\frac{1}{2}v_x+(v+\alpha^{\hat{1}})\partial_x+i\beta\right)\psi=0 \;.
\ee
The Gullstrand-Painlev\'e form~\eqref{dsGP} is ideally suited for studies of analogue black holes~\cite{Unruh:1980cg,Unruh:1994je,Barcelo:2005fc,Schutzhold:2025qna}.
Metrics motivated by such systems typically possess both an outer black-hole horizon and an inner white-hole horizon. The latter generally leads to large blue shifts, which may be both phenomenologically undesirable and numerically demanding. For this reason, we focus on the metric~\eqref{inflation2D} in the examples below.

\section{Scattered-wave-function approach}

We will solve the Dirac equation~\eqref{Dirac1} by generalizing the scattered-wave-function (SWF) approach in~\cite{Torgrimsson:2025pao,Torgrimsson:2025ihd} from QED to GR. We assume that the spacetime is flat in the exterior, but, because of the above discussion, we do not assume that $g_{\mu\nu}\to\eta_{\mu\nu}$. 

\subsection{In and out states}

To define the in and out states, we would like to choose coordinates so that the metric is Minkowski at $t\to\pm\infty$.
Since~\eqref{yaInt} in general leads to multivalued coordinates $y^a$ in 2D, we may not be able to do so globally. However, we can choose one set of coordinates $\hat{y}^a$ to simplify the out states, and another set of coordinates $\check{y}^a$ to simplify the in states. We define the out coordinates $\hat{y}^a$ by replacing $e_\mu^a(t,x)\to e_\mu^a(\infty,{\bf x}):=\hat{e}_\mu({\bf x})$ in~\eqref{yaInt}, and then, since $\partial_\mu\hat{e}_\nu^a=\partial_\nu\hat{e}_\mu^a$ in both the exterior and the interior, the integral is globally path independent and we can for simplicity choose to integrate along straight lines,
\be\label{yhat1}
\hat{y}^a(t,{\bf x})=\int_0^1\ud s\, x^\mu\hat{e}_\mu^a(sx) \;.
\ee

We then define the particle and antiparticle out states as solutions of the full Dirac equation with asymptotic behavior
\be\label{outStates}
\begin{split}
\lim_{t\to+\infty}U_{\rm out}(s,{\bf p},x)&=u_s({\bf p})e^{-ip\hat{y}(x)}\\
\lim_{t\to+\infty}V_{\rm out}(s,{\bf q},x)&=v_s({\bf q})e^{iq\hat{y}(x)} \;.
\end{split}
\ee
The free spinors $u$ and $v$ satisfy  
\be\label{freeSpinorEq}
(\gamma^a p_a-1)u({\bf p})=(\gamma^a q_a+1)v({\bf q})=0 \;,
\ee
and $u^\dagger_r({\bf p})v_s(-{\bf p})=0$ and are normalized as
\be\label{spinorNorm}
u^\dagger_r({\bf p})u_s({\bf p})=
v^\dagger_r({\bf q})v_s({\bf q})=\delta_{rs} \;.
\ee

We similarly define in states as
\be\label{inStates}
\begin{split}
\lim_{t\to-\infty}U_{\rm in}(s,{\bf p},x)&=u_s({\bf p})e^{-ip\check{y}(x)}\\
\lim_{t\to-\infty}V_{\rm in}(s,{\bf q},x)&=v_s({\bf q})e^{iq\check{y}(x)} \;,
\end{split}
\ee
where $\check{y}^a$ are coordinates based on $\check{e}_\mu^a({\bf x})=e_\mu^a(-\infty,{\bf x})$,
\be
\check{y}^a(t,{\bf x})=\int_0^1\ud s\, x^\mu\check{e}_\mu^a(sx)  \;.
\ee

In~\eqref{outStates} we have assumed that the gauge potential $A_\mu\to0$ as $t\to+\infty$, and in~\eqref{inStates} we have assumed that $A_\mu\to0$ as $t\to-\infty$. In 2D we could have $A_\mu\ne0$ asymptotically even though $F_{\mu\nu}\to0$. This leads to a problem similar to the one discussed above, which was dealt with in~\cite{Torgrimsson:2025pao}. In this paper, we assume for simplicity that $A_\mu\to0$ asymptotically.

\subsection{Inner product}

To obtain the pair production amplitude, we need an inner product between the in and out states. Current conservation,
\be
\partial_\mu(\sqrt{g}j^\mu)=0
\qquad
j^\mu=\bar{\psi}\gamma^\mu\psi
\qquad
\bar{\psi}=\psi^\dagger\gamma^{\hat{0}} \;,
\ee
implies a time-independent inner product as~\cite{Parker:1980kw,Parker:1980hlc}
\be\label{innerg}
\langle\psi|\varphi\rangle=\int\ud^3{\bf x}\sqrt{g}\psi^\dagger\gamma^{\hat{0}}\gamma^{0}\varphi \;.
\ee
In the time gauge~\eqref{timeGauge} we have
\be
\sqrt{g}=e^{\hat{0}}_0\det e^{\hat{k}}_j
\qquad
\gamma^{\hat{0}}\gamma^0=\frac{1}{e^{\hat{0}}_0} \;,
\ee
so the inner product simplifies to
\be\label{innerADM}
\langle\psi|\varphi\rangle=\int\ud^3{\bf x}\,\text{det }e^{\hat{i}}_j\,\psi^\dagger\varphi \;.
\ee

The expectation value of particles or antiparticles in the state $m=(s,{\bf p})$ is given by the well-known formulas~\cite{unstableVacuumBook}
\be\label{Nem}
\begin{split}
N_{e^\LCm}(m)&={}_{\rm in}\langle0|a_{\rm out}^\dagger(m)a_{\rm out}(m)|0\rangle_{\rm in}\\
&=[\langle U_{\rm out}|V_{\rm in}\rangle\langle V_{\rm in}|U_{\rm out}\rangle]_{mm} \\
N_{e^\LCp}(m)&={}_{\rm in}\langle0|b_{\rm out}^\dagger(m)b_{\rm out}(m)|0\rangle_{\rm in}\\
&=[\langle V_{\rm out}|U_{\rm in}\rangle\langle U_{\rm in}|V_{\rm out}\rangle]_{mm} \;,
\end{split}
\ee
where a sum over the in states is implied,
\be
|U_{\rm in}\rangle\langle U_{\rm in}|=\int\frac{\ud^3{\bf k}}{(2\pi)^3}\sum_{w}|U_{\rm in}(w,{\bf k})\rangle\langle U_{\rm in}(w,{\bf k})| \;.
\ee
These formulas do not give any information about the correlation between the particle and antiparticle. Formulas for the correlation were derived in~\cite{Torgrimsson:2025pao}, 
\be
\begin{split}
N_{e^\LCm e^\LCp}(m,n)&={}_{\rm in}\langle0|a_{\rm out}^\dagger(m)a_{\rm out}(m)b_{\rm out}^\dagger(n)b_{\rm out}(n)|0\rangle_{\rm in}\\
&=N_{e^\LCm}(m)N_{e^\LCp}(n)+N_1(m,n) \;,
\end{split}
\ee
where $N_1$ encodes the particle-antiparticle correlation
\be\label{N1}
\begin{split}
N_1&=|{}_m\langle U_{\rm out}|U_{\rm in}\rangle\langle U_{\rm in}|V_{\rm out}\rangle_n|^2\\
&=|{}_m\langle U_{\rm out}|V_{\rm in}\rangle\langle V_{\rm in}|V_{\rm out}\rangle_n|^2\;.
\end{split}
\ee

Since the inner product is time independent, we can evaluate $\langle U_{\rm out}(s,{\bf p})|U_{\rm out}(r,{\bf q})\rangle$ at $t\to+\infty$, where the integrability condition $\partial_\mu e^{\hat{0}}_\nu=\partial_\nu e^{\hat{0}}_\mu$ implies $\partial_i e^{\hat{0}}_0=0$ and hence $y^{\hat{0}}(t)$ is independent of $x^i$. Therefore, changing integration variables from $x^i$ to $y^{\hat{i}}$ leads to a Jacobian that cancels $\text{det }e^{\hat{i}}_j$ and gives
\be
e^{i(p_{\hat{0}}-q_{\hat{0}})y^{\hat{0}}}\int\ud^3{\bf y} e^{i(p-q)_{\hat{i}}y^{\hat{i}}}=(2\pi)^3\delta^3({\bf p}-{\bf q}) \;.
\ee
Thus, with the free spinors normalized as~\eqref{spinorNorm} we have
\be
\begin{split}
\langle U_{\rm out}(s,{\bf p})|U_{\rm out}(r,{\bf q})\rangle&=(2\pi)^3\delta^3({\bf p}-{\bf q})\delta_{sr}\\
\langle V_{\rm out}(s,{\bf p})|V_{\rm out}(r,{\bf q})\rangle&=(2\pi)^3\delta^3({\bf p}-{\bf q})\delta_{sr}\\
\langle U_{\rm out}(s,{\bf p})|V_{\rm out}(r,{\bf q})\rangle&=0 
\end{split}
\ee
and similarly for $U_{\rm out},V_{\rm out}\to U_{\rm in},V_{\rm in}$.

\subsection{Background and scattered waves}

We solve the Dirac equation by splitting the wave function into a background and a scattered wave,
\be
\psi_{\rm out}=\psi_{\rm back}+\psi_{\rm scat} \;,
\ee
where the background waves are given by the future-asymptotic form of the out states,
\be\label{backy}
U_{\rm back}=ue^{-ip\hat{y}} \quad
V_{\rm back}=ve^{iq\hat{y}} \;,
\ee
and the scattered waves are obtained by numerically solving
\be\label{scatDirac}
\begin{split}
(i\slashed{D}-1)\psi_{\rm scat}&=-(i\slashed{D}-1)\psi_{\rm back} \\
\psi_{\rm scat}(t_\LCp,{\bf x})&=0 
\end{split}
\ee
from the asymptotic future to the asymptotic past, i.e. from some $t_\LCp\gg0$ to some $t_\LCm\ll0$, chosen such that the spacetime is approximately flat and $A_\mu\approx0$ for $t>t_\LCp$ and $t<t_\LCm$. 

With the Dirac equation expressed as in~\eqref{dt_Dirac}, the background wave obeys
\be\label{dt_Dirac_back}
\partial_t\psi_{\rm back}=-\left[-i\epsilon\hat{p}_k\hat{\alpha}^k+\hat{W}\right]\psi_{\rm back} \;,
\ee
where $\hat{\alpha}^k({\bf x})=\alpha^k(\infty,{\bf x})$, $\hat{W}({\bf x})=W(\infty,{\bf x})$ and
\be
\hat{p}_\mu({\bf x})=\hat{e}^a_\mu({\bf x})p_a \;,
\ee
with $\epsilon=+1$ for $U_{\rm back}({\bf p})$ and $\epsilon=-1$ for $V_{\rm back}({\bf p})$. Subtracting~\eqref{dt_Dirac_back} from~\eqref{dt_Dirac} gives
\be\label{dt_Dirac_SWF}
\begin{split}
\partial_t\psi_{\rm scat}=&-\left[iA_0+\alpha^k(\partial_k+iA_k)+W\right]\psi_{\rm scat}\\
&-\left[iA_0+\alpha^k iA_k-i\epsilon\hat{p}_k\Delta\alpha^k+\Delta W\right]\psi_{\rm back}\;,
\end{split}
\ee
where $\Delta\alpha^k=\alpha^k-\hat{\alpha}^k$ and $\Delta W=W-\hat{W}$. Eq.~\eqref{dt_Dirac_SWF} is numerically convenient because all the terms vanish outside a backward lightcone, i.e. for $|{\bf x}|>L(t)$ with $L(t_\LCm)>L(t)>L(t_\LCp)\approx0$.

For the example in~\eqref{inflation2D} we have
\be
\begin{split}
&\partial_t\psi_{\rm scat}=
-\left(\frac{1}{2}\partial_t A+\alpha^{\hat{1}} e^{-A}\partial_x+i\beta\right)\psi_{\rm scat} \\
-&\left(\frac{1}{2}\partial_t A-i\epsilon p_{\hat{1}}\alpha^{\hat{1}}\left[e^{-A(t,x)+A(\infty,x)}-1\right]\right)\psi_{\rm back}\;.
\end{split}
\ee

For spacetimes with trivial holonomy, it is possible to choose coordinates so that $g_{\mu\nu}\to\eta_{\mu\nu}$ everywhere in the exterior region, and then
$\hat{y}^a=\check{y}^a$ so $U_{\rm back}(t_\LCm)=U_{\rm in}(t_\LCm)$,
\be\label{deltaSeparated}
\begin{split}
\langle U_{\rm in}(w,{\bf k})|U_{\rm back}(s,{\bf p})\rangle&=(2\pi)^3\delta({\bf p}-{\bf k})\delta_{sw}\\
\langle U_{\rm in}|V_{\rm back}\rangle&=0
\end{split}
\ee
and the momentum correlation~\eqref{N1} becomes
\be\label{Nmn}
\begin{split}
N_1=\big|&\langle U_{\rm back}(s,{\bf p})|V_{\rm scat}(r,{\bf q})\rangle\\
+&\langle U_{\rm scat}(s,{\bf p})|U_{\rm in}\rangle\langle U_{\rm in}|V_{\rm scat}(r,{\bf q})\rangle\big|^2 \;,
\end{split}
\ee
where the inner products are given by~\eqref{innerADM} evaluated at $t_\LCm$. Since all of these inner products involve either $U_{\rm scat}$ or $V_{\rm scat}$, both of which are negligible outside the backward lightcone, i.e. for $|{\bf x}|>L(t_\LCm)$, the integrals in~\eqref{Nmn} receive contributions only from the finite region $|{\bf x}|<L(t_\LCm)$. Consequently, they cannot generate delta functions, and all ingredients entering~\eqref{Nmn} are numerically well behaved.  
For 2D metrics with nontrivial holonomy, obtaining similarly numerically well-behaved formulas requires additional work; two methods are presented in Appendix~\ref{twoMethods}.

\subsection{Numerical examples}

\begin{figure*}
    \centering
    \includegraphics[width=.49\linewidth]{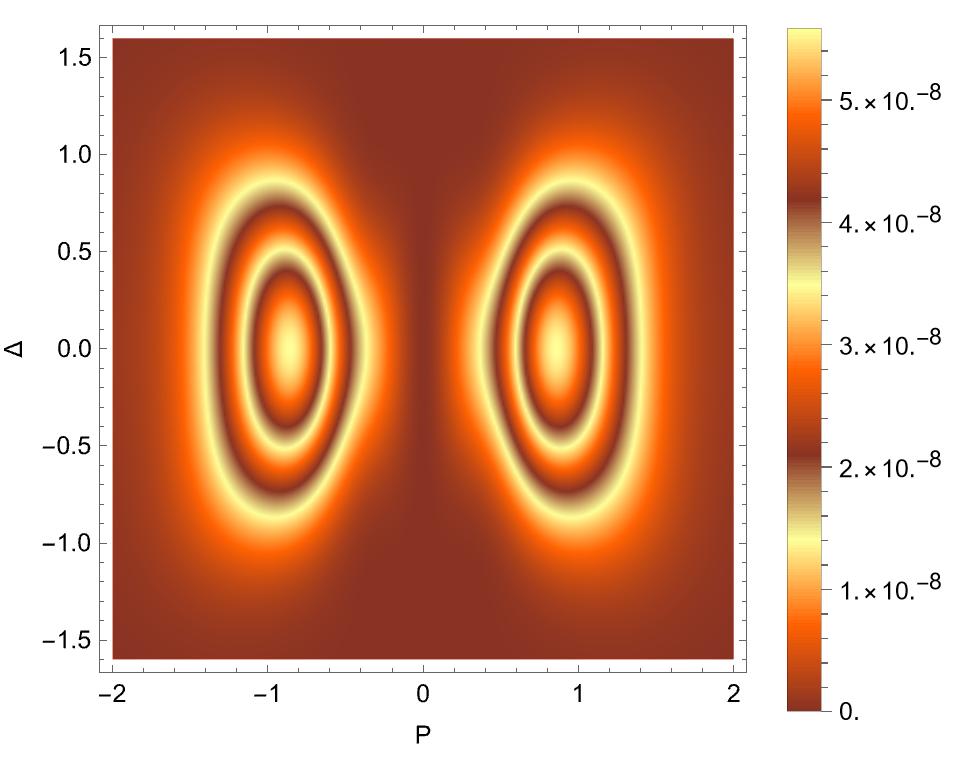}
    \includegraphics[width=.49\linewidth]{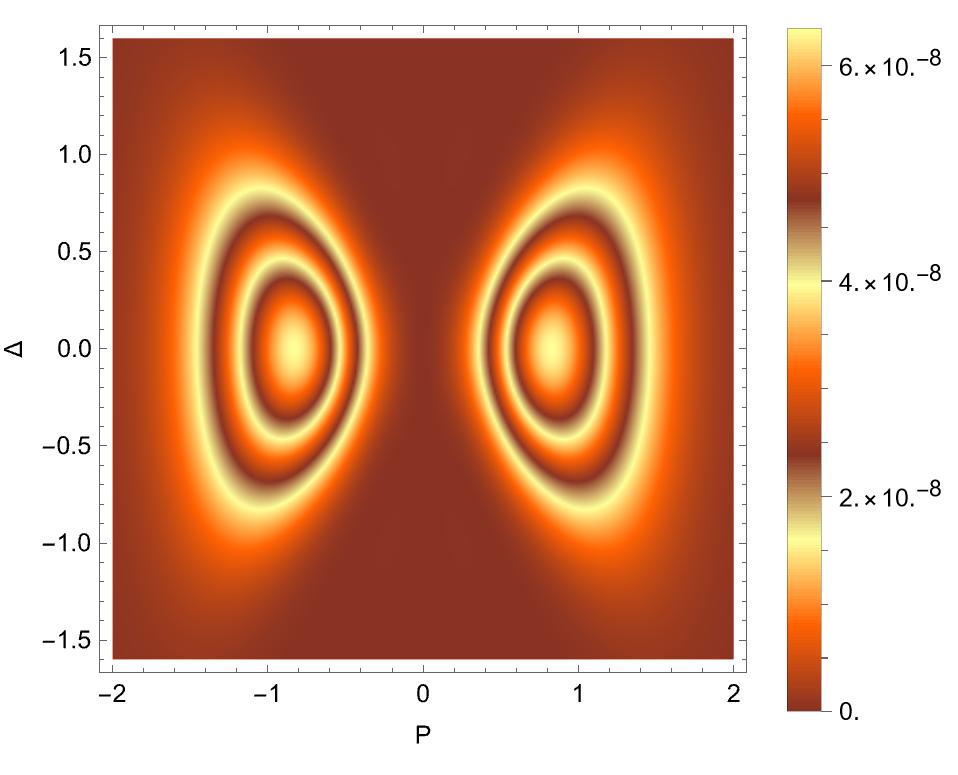}
    \caption{Spectrum of pair production with $P=(q^1-p^1)/2$ and $\Delta=q^1+p^1$, where $p$ and $q$ are the particle and antiparticle momenta, for the metric in~\eqref{Agauss} with $\alpha=1$ and $\omega=\kappa=0.3$. The first plot shows the quadratic-instanton approximation and the second plot the exact SWF result.}
    \label{fig:gauss_2D_03}
\end{figure*}

\begin{figure}
    \centering
    \includegraphics[width=\linewidth]{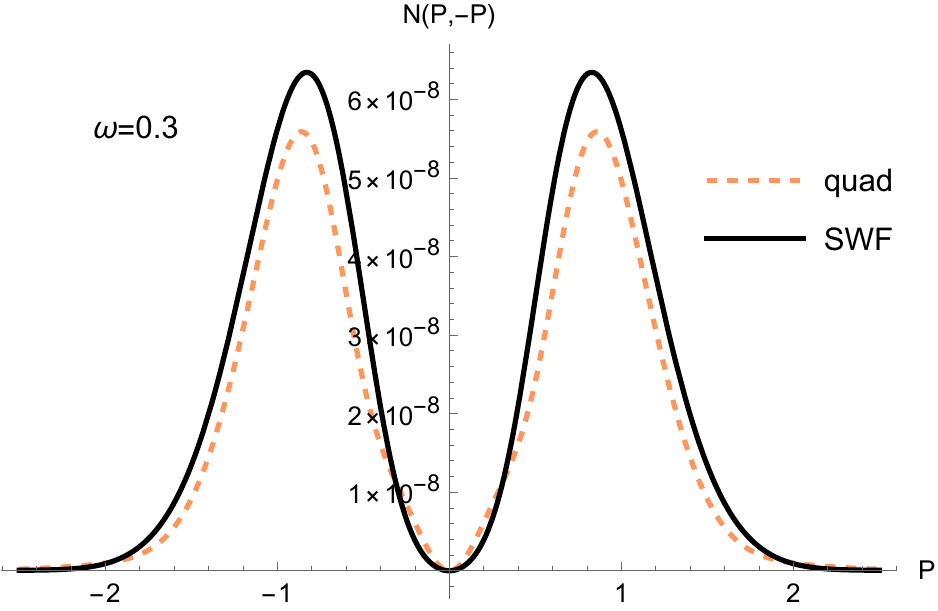}
    \caption{The line $\Delta=0$ in Fig.~\eqref{fig:gauss_2D_03}.}
    \label{fig:gauss_1D_03}
\end{figure}

\begin{figure*}
    \centering
    \includegraphics[width=.49\linewidth]{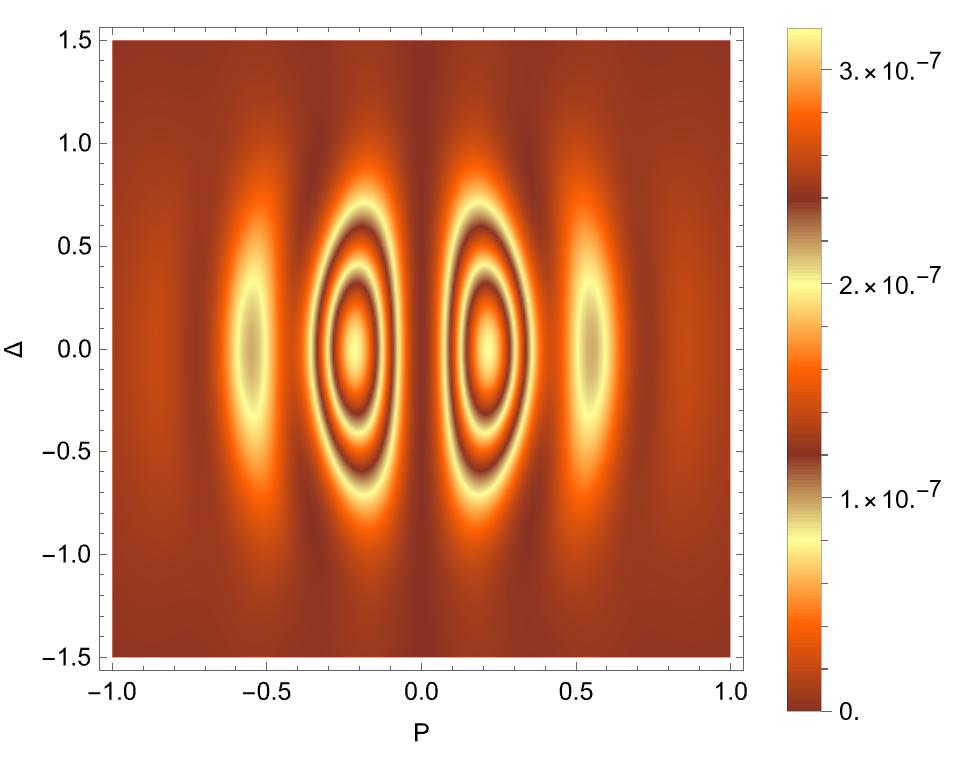}
    \includegraphics[width=.49\linewidth]{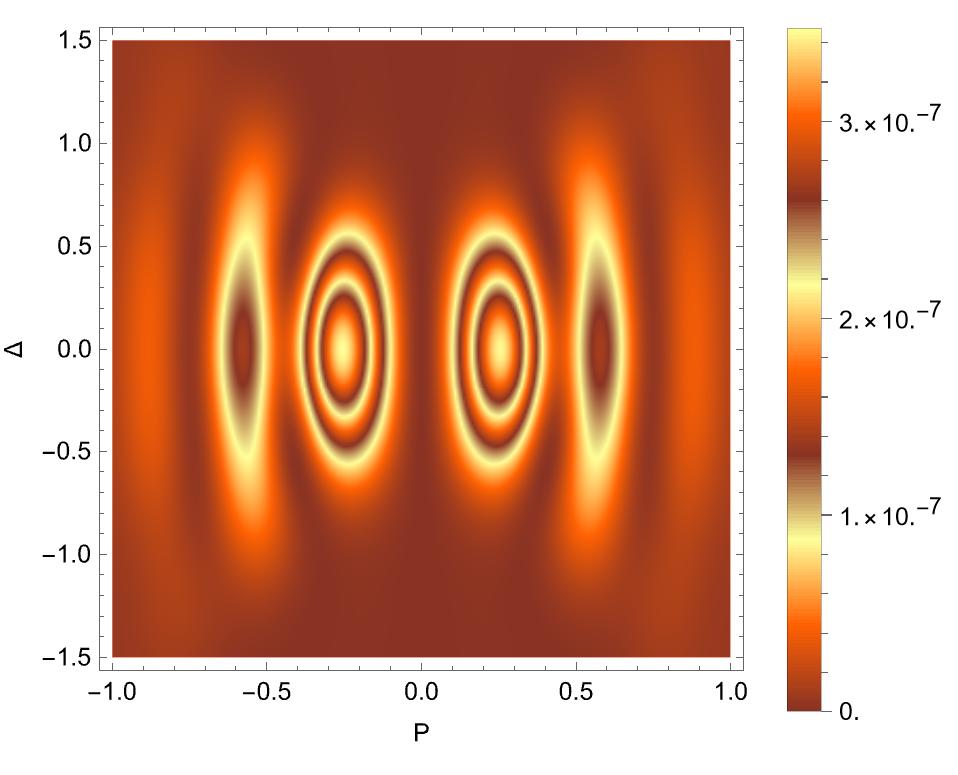}
    \caption{Same as Fig.~\ref{fig:gauss_2D_03} but with $\alpha=-2$. Note that the colors are scaled relative to the maximum value in each plot.}
    \label{fig:minus_gauss_2D_03}
\end{figure*}

\begin{figure}
    \centering
    \includegraphics[width=\linewidth]{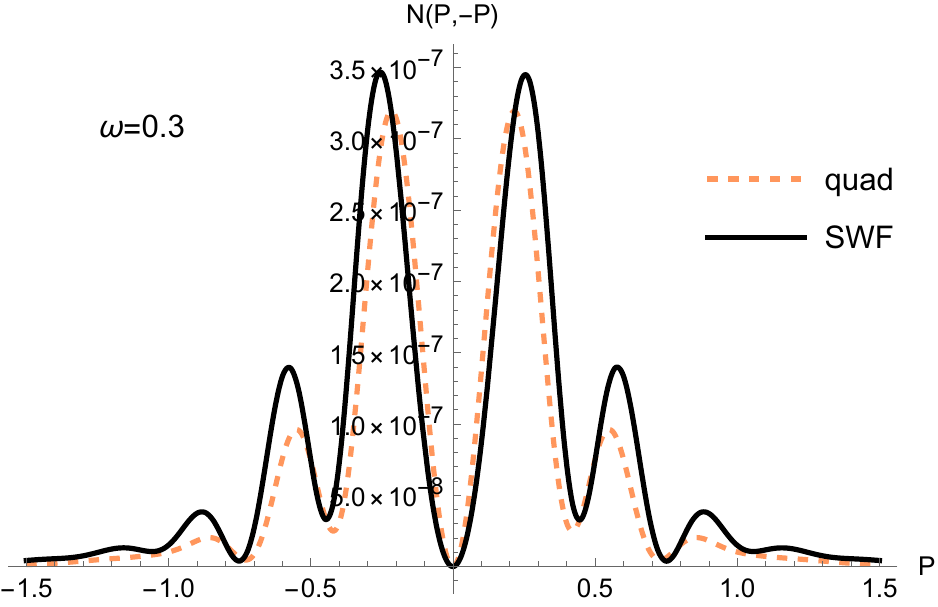}
    \caption{The line $\Delta=0$ in Fig.~\eqref{fig:minus_gauss_2D_03}.}
    \label{fig:minus_gauss_1D_03}
\end{figure}

\begin{figure}
    \centering
    \includegraphics[width=.8\linewidth]{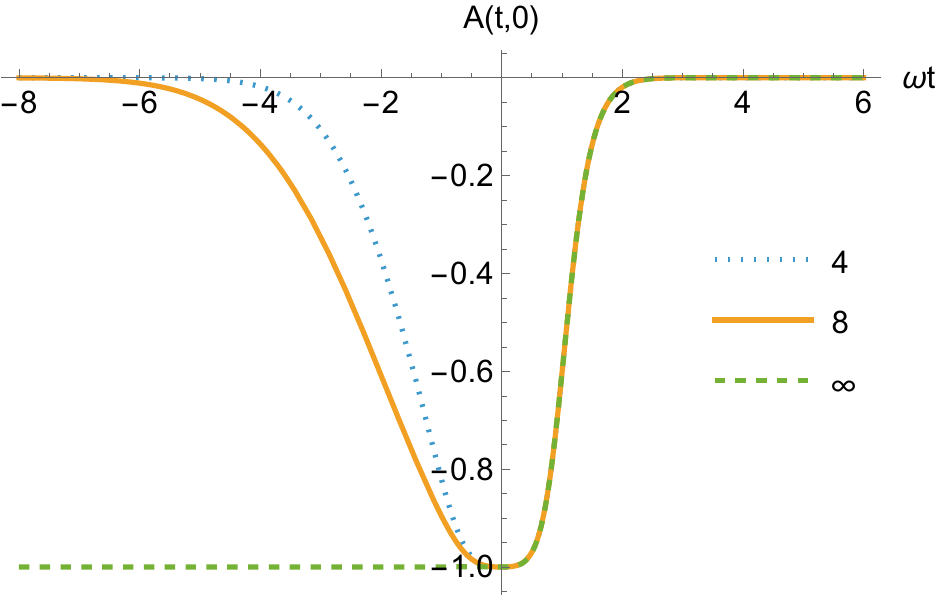}\\
    \includegraphics[width=.8\linewidth]{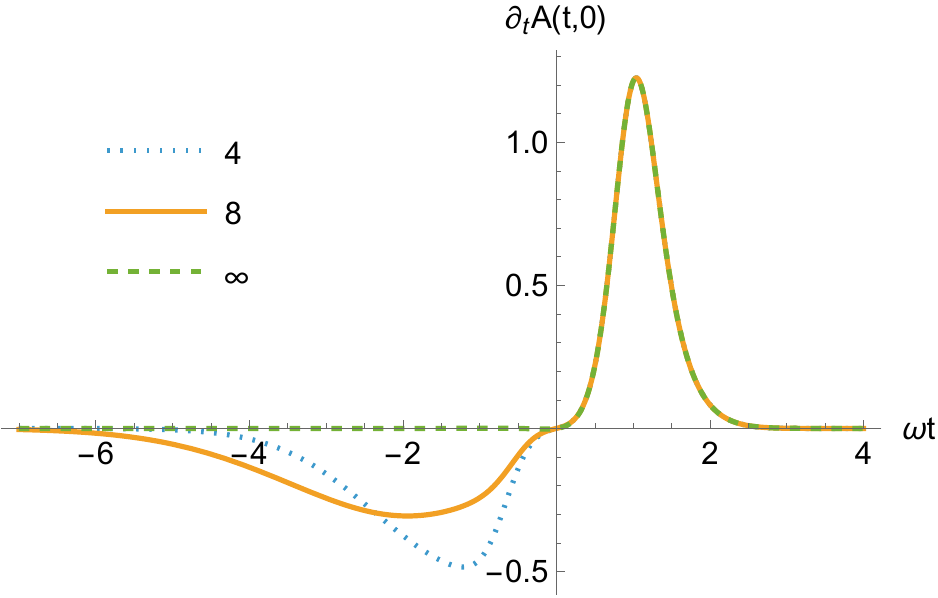}
    \caption{The function in~\eqref{Ad} with $\alpha=-1$ and $d=4,8,\infty$.}
    \label{fig:dA}
\end{figure}

\begin{figure}
    \centering
    \includegraphics[width=\linewidth]{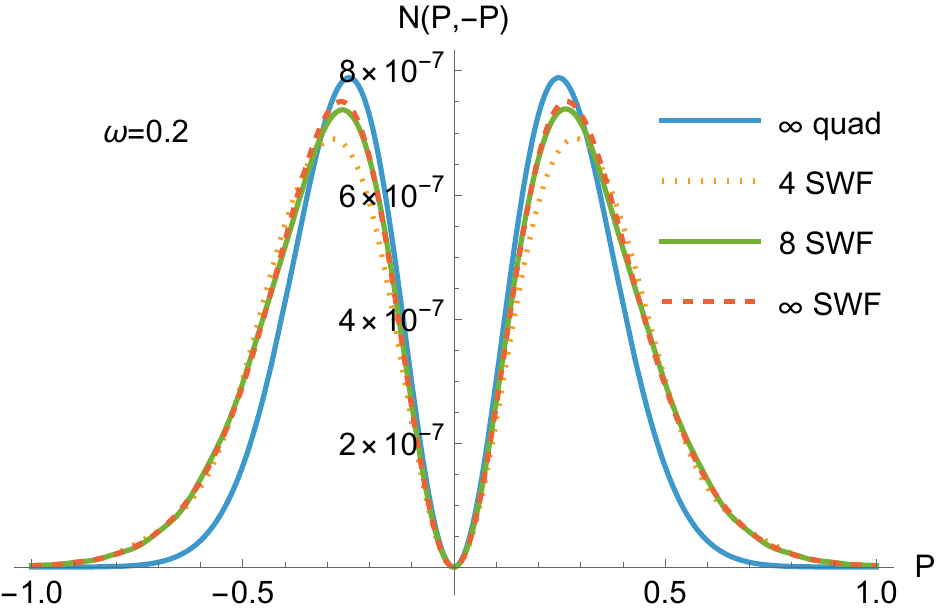}
    \caption{The spectrum for the metrics in Fig.~\ref{fig:dA}.}
    \label{fig:large_d}
\end{figure}

\begin{figure*}
    \centering
    \includegraphics[width=.49\linewidth]{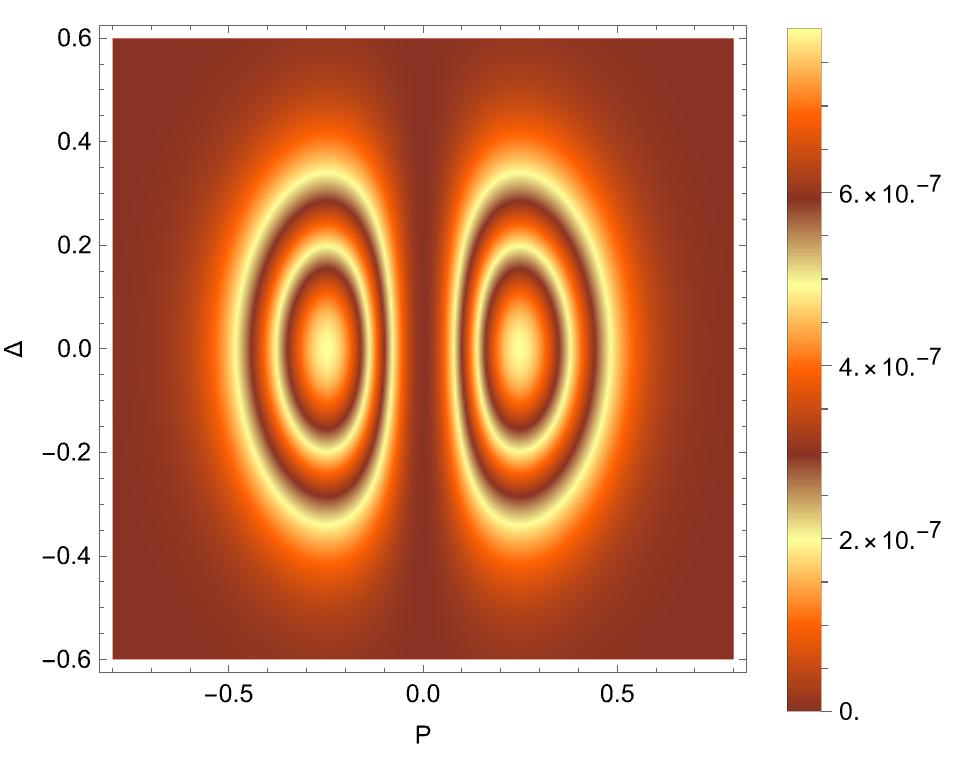}
    \includegraphics[width=.49\linewidth]{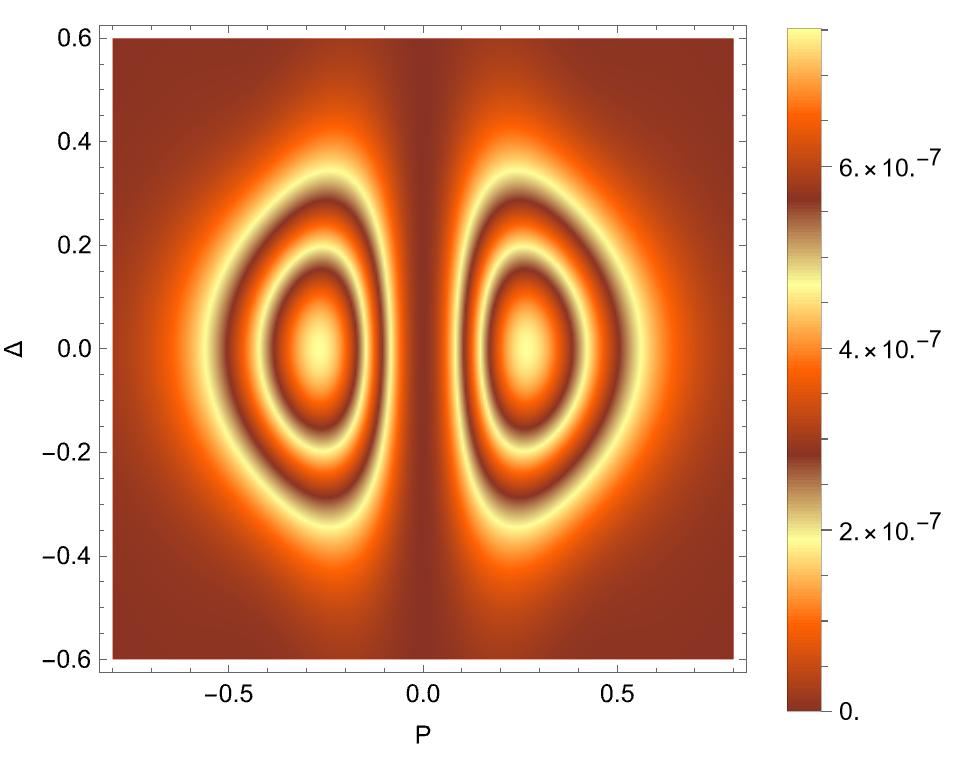}
    \caption{Similar to Fig.~\ref{fig:gauss_2D_03} but for the field in~\eqref{Ad} with $d=\infty$, which gives a nontrivial holonomy.}
    \label{fig:holonomy2D}
\end{figure*}

We solve~\eqref{dt_Dirac_SWF} by discretizing the spatial coordinates, computing the spatial derivatives with FFT, and evolving in time with an adaptive step-size solver. This can be conveniently implemented in Python using the JAX library~\cite{jax}, which allows for GPU acceleration with a simple NumPy interface, and the Diffrax solver~\cite{diffrax}. This approach is described in detail in~\cite{Torgrimsson:2025pao,Torgrimsson:2025ihd} for the QED case, and Jupyter notebooks can be found at~\cite{GTGitHub}. 

In the plots below, we use ${\bf p}$ and ${\bf q}$ for the particle and antiparticle momenta, or $P_i$ and $\Delta_i$ defined as
\be
p_i=-P_i+\frac{\Delta_i}{2} \qquad q_i=P_i+\frac{\Delta_i}{2} \;.
\ee
A space-independent field would give $\delta^3({\bf \Delta})$. 

As the first example, we consider a metric of the form~\eqref{inflation2D} with
\be\label{Agauss}
A(t,x)=\alpha e^{-(\omega t)^2-(\kappa x)^2} \;,
\ee
which describes a spacetime with both inflation and deflation and with a trivial holonomy.
Figs.~\ref{fig:gauss_2D_03} and~\ref{fig:gauss_1D_03} show the results for $\alpha=1$ and $\omega=\kappa=0.3$. Figs.~\ref{fig:minus_gauss_2D_03} and~\ref{fig:minus_gauss_1D_03} show results for $\alpha=-2$ and $\omega=\kappa=0.3$. We find good agreement between the exact SWF results and the quadratic-instanton approximation, especially given that the small parameter that justifies the instanton approximation, $\omega=0.3$, is not particularly small and one should in general expect a relative error of $\mathcal{O}(\omega)$.

We next consider
\be\label{Ad}
A(t,x)=\alpha\exp\left(-[D(t)+I(t)](\omega t)^2-(\kappa x)^2\right) \;,
\ee
where
\be
D(t)=\frac{1}{d+e^{4(\omega t+1)}}
\qquad
I(t)=\frac{1}{1+e^{4(\omega t-1)}} \;.
\ee
With $\alpha<0$, this metric describes a spacetime with a period of deflation controlled by $D(t)$, and subsequently a period of inflation controlled by $I(t)$. As illustrated in Fig.~\ref{fig:dA}, by increasing $d$ we obtain a metric with an increasingly slow deflation, without changing the inflation. For $d\gg1$ there will be a wide region around $t=0$ where the metric $g_{\mu\nu}(d)$ looks like $g_{\mu\nu}(\infty)$. But $\lim_{t\to-\infty}g_{\mu\nu}=0$ for any finite value of $d$, while $\lim_{t\to-\infty}\lim_{d\to\infty}g_{\mu\nu}\ne0$, so these two limits do not commute. One may therefore wonder whether the probability $P(d=\infty)$ might be qualitatively different from $P(1\ll d\ne\infty)$, but Fig.~\ref{fig:large_d} shows that $P(1\ll d\ne\infty)\approx P(d=\infty)$. This can be naturally understood in the instanton formalism. 

There are in general multiple instantons. In its creation region, an instanton $x^\mu(u)$ is complex, but its real part is close to a region where the curvature has a local maximum. As seen in Fig.~\ref{fig:dA}, there is one peak for the deflation and one for the inflation. As $d$ increases, the deflation peak shrinks and eventually disappears as $d\to\infty$. As the height of the peak effectively controls the amount of exponential suppression, and since even a factor of $2$ can mean a dramatic change of the exponential $\exp(-\mathcal{O}(1)/[\omega/2])\ll\exp(-\mathcal{O}(1)/\omega)$ for $\omega\ll1$, the contribution from the deflation peak becomes negligible already for moderately large $d$. Since both the particle and the antiparticle ends of the instanton move toward the asymptotic future, $t(u)\to\infty$ as either $\text{Re }u\to+\infty$ or $\text{Re }u\to-\infty$, the instantons created by the inflation peak never enter the $t\ll0$ region and are therefore insensitive to whether $A(t=-\infty,x)$ is zero or not, as long as the deflation is slow enough.     

Thus, the convergence to the $d=\infty$ result is expected. However, while the instanton approach is much simpler for $d=\infty$ compared to $d=1$ (because there are fewer instantons to consider for $d=\infty$), the SWF approach is simpler for $d=1$. In order to use~\eqref{Nmn} we need to integrate the Dirac equation from $t_\LCp$ backwards in time to $t_\LCm$, where $t_\LCm$ must be chosen such that $g_{\mu\nu}(t<t_\LCm)\approx\eta_{\mu\nu}$. But if the deflation period is slow, then $|t_\LCm|$ has to be large, which can mean many more time steps and hence a slow code. However, for $d=\infty$ we can instead use the methods presented in Appendix~\ref{twoMethods}, which work also for metrics with a nontrivial holonomy, where $t_\LCm$, chosen such that $g_{\mu\nu}(t<t_\LCm,x)\approx g_{\mu\nu}(-\infty,x)$, does not have to be very large. The $d=4,8$ results in Fig.~\ref{fig:large_d} have been obtained using~\eqref{Nmn}, while the $d=\infty$ (SWF) result has been obtained using either of the two methods in Appendix~\ref{twoMethods}. The fact that the $d=8$ and $d=\infty$ results are close is a nontrivial check of these three different SWF methods. 

Finally, Fig.~\ref{fig:holonomy2D} shows the 2D spectra for $d=\infty$ obtained using the quadratic instanton approximation and the SWF approach.

\subsection{Exchange antisymmetry}

In the examples above, the pair-production spectrum vanishes when the particle and antiparticle have identical momenta. This can be traced to the fact that gravity does not distinguish particles from antiparticles, while the underlying field remains fermionic. The resulting suppression is reminiscent of the Pauli exclusion principle. We can demonstrate this as follows.

The amplitude in~\eqref{N1} can be expressed in two ways
\be\label{N1amp}
\begin{split}
M(m,n)&={}_m\langle U_{\rm out}|U_{\rm in}\rangle\langle U_{\rm in}|V_{\rm out}\rangle_n\\
&=-{}_m\langle U_{\rm out}|V_{\rm in}\rangle\langle V_{\rm in}|V_{\rm out}\rangle_n \;,
\end{split}
\ee
where we used the completeness relation
\be
|U_{\rm in}\rangle\langle U_{\rm in}|+|V_{\rm in}\rangle\langle V_{\rm in}|=1
\ee
together with ${}_m\langle U_{\rm out}|V_{\rm out}\rangle_n=0$. In a Majorana representation of $\gamma^a$, $\text{Re }\gamma^a=0$, and in the absence of any EM field, $A_\mu=0$, the antiparticle solutions to the Dirac equation are simply the complex conjugate of the particle solutions, $V_{\rm in/out}(s,{\bf p},x)=U_{\rm in/out}^*(s,{\bf p},x)$. Substituting this into~\eqref{N1amp} shows that the amplitude is antisymmetric under exchange of the particle and antiparticle quantum numbers, $M(m,n)=-M(n,m)$, and consequently vanishes when these quantum numbers coincide.

In the 2D examples above, the spin states decouple into two independent sectors with either $s=r=-1$ (particle with spin down and antiparticle with spin up) or $s=r=+1$ (particle with spin up and antiparticle with spin down). Within each sector, the spectrum therefore vanishes along the diagonal $p=q$ ($P=q-p=0$).

\section{Derivation of path integral}

In the next section we develop a worldline instanton formalism that provides a semiclassical approximation to the exact SWF result described in the previous section. Before turning to the instanton approximation, we first derive an exact worldline-path-integral representation for the fermion propagator in a general curved spacetime. Such worldline integrals have a long history; see~\cite{Schubert:2001he,Bastianelli:2006rx,Edwards:2019eby} for reviews. They almost always (as far as we are aware) involve Grassmann path integrals for the spin degrees of freedom, as well as ghost integrals arising from exponentiating factors of $\prod_{0<\tau<1}\sqrt{g}$ in the path-integration measure $\mathcal{D}x$. While such representations are convenient for many applications, for instanton calculations it is more useful to work with path-ordered exponentials of gamma matrices, since these become straightforward to evaluate on the instanton solutions. In this section we present a detailed derivation of such a representation. Although most steps are standard, none of the references we are aware of uses exactly the same conventions, so we reproduce the derivation here in order to fix signs and normalizations.   

\subsection{Squaring the Dirac operator}

The goal in this section is to derive a worldline representation of the propagator
\be
(i\slashed{D}_x-1)G(x,y)=\frac{1}{\sqrt{g}}\delta^4(x-y) \;,
\ee
which can then be used in a LSZ-reduction approach to obtain the pair-production amplitude.
The first step is to Klein-Gordonize the Dirac operator by writing 
\be\label{GtildeG}
G(x,y)=(i\slashed{D}_x+1)\tilde{G}(x,y) \;,
\ee
so that $\tilde{G}$ obeys an equation,
\be\label{Kx}
-K_x\tilde{G}(x,y)=\frac{1}{\sqrt{g}}\delta^4(x-y) \;,
\ee
with a squared Dirac operator,
\be
K=-(i\slashed{D}-1)(i\slashed{D}+1)=\gamma^\mu D_\mu\gamma^\nu D_\nu+1 \;.
\ee
In the literature, the symbol $D_\mu$ is often used with different meanings depending on what object it acts on. In this section, however, $D_\mu$ is always defined precisely by~\eqref{covariantDer}. In particular,
\be
[D_\mu,\gamma^\nu]=-\Gamma^\nu_{\mu\rho}\gamma^\rho 
\ee
and using $\gamma^\mu\gamma^\nu=g^{\mu\nu}-i\sigma^{\mu\nu}$ we obtain 
\be
K=g^{\mu\nu}(D_\mu D_\nu-\Gamma^\rho_{\mu\nu}D_\rho)-\frac{i}{2}\sigma^{\mu\nu}[D_\mu,D_\nu]+1 \;.
\ee
To simplify the commutator, we first calculate
\be
\begin{split}
\partial_\mu\omega_{\nu ab}-(\mu\leftrightarrow\nu)&=e_a^\rho e_b^\sigma R_{\rho\sigma\mu\nu}\\
&-({\omega_{\mu a}}^c\omega_{\nu cb}-{\omega_{\nu a}}^c\omega_{\mu cb}) \;.
\end{split}
\ee
The $\omega\omega$ terms cancel against the terms coming from $[\omega\sigma,\omega\sigma]$, so
\be
[D_\mu,D_\nu]=\frac{1}{4}R_{\mu\nu\rho\sigma}\gamma^\rho\gamma^\sigma
+iF_{\mu\nu}\;.
\ee
From the cyclic identity of the Riemann tensor we have
\be
\begin{split}
0&=(R_{\mu\nu\rho\sigma}+R_{\mu\rho\sigma\nu}+R_{\mu\sigma\nu\rho})\gamma^\mu\gamma^\nu\gamma^\rho\gamma^\sigma\\
&=R_{\mu\nu\rho\sigma}(\gamma^\mu\gamma^\nu\gamma^\rho\gamma^\sigma+\gamma^\mu\gamma^\sigma\gamma^\nu\gamma^\rho+\gamma^\mu\gamma^\rho\gamma^\sigma\gamma^\nu)\\
&=3R_{\mu\nu\rho\sigma}(\gamma^\mu\gamma^\nu\gamma^\rho\gamma^\sigma+2g^{\nu\sigma}\gamma^\mu\gamma^\rho) \;,
\end{split}
\ee
where in the third step we rewrote $\gamma^\sigma\gamma^\nu\gamma^\rho=\gamma^\nu\gamma^\rho\gamma^\sigma+\dots$ and $\gamma^\rho\gamma^\sigma\gamma^\nu=\gamma^\nu\gamma^\rho\gamma^\sigma+\dots$ and used $R_{\mu\nu\rho\sigma}=-R_{\mu\nu\sigma\rho}$. With $R_{\mu\nu\rho\sigma}=-R_{\nu\mu\rho\sigma}$ and $R_{\mu\nu}=R_{\nu\mu}=R^\rho_{\mu\nu\rho}$ we then find   
\be
R_{\mu\nu\rho\sigma}\gamma^\mu\gamma^\nu\gamma^\rho\gamma^\sigma=2R_{\mu\nu}\gamma^\mu\gamma^\nu=2R \;.
\ee
Thus, the squared Dirac operator becomes
\be\label{KbeforeWeyl}
K=g^{\mu\nu}(D_\mu D_\nu-\Gamma^\rho_{\mu\nu}D_\rho)+\frac{R}{4}+\frac{1}{2}\sigma^{\mu\nu}F_{\mu\nu}+1 \;.
\ee

\subsection{Schwinger proper time integral}

We introduce operators $\hat{x}^\mu$ and $\hat{p}_\nu$ that satisfy~\cite{Schwinger:1951nm}
\be
[\hat{x}^\mu,\hat{p}_\nu]=i\delta^\mu_\nu \;.
\ee
One could choose $\hat{p}_\mu=-i\partial_\mu$~\cite{deAlfaro:1987bu}, but it is more common to choose~\cite{Grosche:1992qx,Bastianelli:2006rx}
\be\label{pOp1}
\hat{p}_\mu=-\frac{1}{g^{1/4}}i\partial_\mu g^{1/4} 
=-i\left(\partial_\mu+\frac{1}{2}\Gamma^\rho_{\rho\mu}\right) \;,
\ee
so that both $\hat{x}^\mu$ and $\hat{p}_\mu$ are hermitian with respect to the inner product
\be
\langle\varphi|\phi\rangle=\int\ud^4x\sqrt{g}\varphi^\dagger\phi \;.
\ee
The equivalence of the two expressions for $\hat{p}_\mu$ follows from
\be\label{gder}
\partial_\mu\ln g=g^{\alpha\beta}g_{\alpha\beta,\mu}
\qquad
\frac{\partial_\mu g^a}{g^a}=a g^{\alpha\beta}g_{\alpha\beta,\mu}=2a\Gamma^\rho_{\rho\mu} \;.
\ee

The eigenstates
\be
\hat{x}^\mu|x\rangle=x^\mu|x\rangle
\qquad
\hat{p}_\mu|p\rangle=p_\mu|p\rangle
\ee
are normalized as
\be
\int\ud^4x\sqrt{g}|x\rangle\langle x|=\int\frac{\ud^4p}{(2\pi)^4}|p\rangle\langle p|=1 \;,
\ee
which implies
\be
\langle x|y\rangle=\frac{1}{\sqrt{g}}\delta^4(x-y)
\qquad
\langle x|p\rangle=\frac{e^{ip_\mu x^\mu}}{g^{1/4}} \;.
\ee

We can view~\eqref{Kx} as the $\langle x|\dots|y\rangle$ matrix element of the operator equation
\be
-K\tilde{G}=1 \;,
\ee
where $K$ is given by~\eqref{KbeforeWeyl} but with $x\to\hat{x}$ and $\partial_\mu\to g^{1/4}i\hat{p}_\mu g^{-1/4}$, and then solve it by introducing a Schwinger proper-time integral,
\be\label{Tint}
\tilde{G}(x,y)=\int_0^\infty\!\frac{\ud T}{2i}\mathcal{G}
\qquad
\mathcal{G}(T,x,y)=\langle x|e^{-\frac{iT}{2}K}|y\rangle
\;.
\ee

\subsection{Feynman path integral}

The next step is to turn~\eqref{Tint} into a Feynman path integral using Dirac's idea of splitting the ``time'' ($T$ in our case) into many smaller steps~\cite{Dirac:1933xn}, i.e. the $N\to\infty$ limit of
\be\label{DiracSteps}
\begin{split}
\mathcal{G}(x,y)=\prod_{i=1}^{N-1}\ud^4 x_i\sqrt{g(x_i)}\prod_{j=1}^N\langle x_j|e^{-\frac{i\delta}{2}K}|x_{j-1}\rangle \;,
\end{split}
\ee
where $\delta=T/N$, $x_0^\mu=y^\mu$ and $x_N^\mu=x^\mu$. A common approach~\cite{Bastianelli:2006rx} for the next step is to Weyl-order and insert $\int\ud^4p_i|p_i\rangle\langle p_i|$ between the $x_i$ integrals. However, given existing results in the literature, it is not too difficult to guess the form of $\langle x_j|e^{-i\delta K/2}|x_{j-1}\rangle$ for $\delta\ll1$. We make the following ansatz
\be\label{matrixAnsatz}
\begin{split}
&\langle x_j|e^{-\frac{i\delta}{2}K}|x_{j-1}\rangle=\frac{1}{(2\pi\delta)^2}\frac{\sqrt{g(\tilde{x}_j)}}{[g(x_j)g(x_{j-1})]^{1/4}}\\
&\times\exp\bigg(-\frac{i\delta}{2}\bigg[g_{\mu\nu}\frac{\Delta x_j^\mu\Delta x_j^\nu}{\delta^2}+2\frac{\Delta x_j^\mu}{\delta}a_\mu+b\bigg]\bigg) \;,
\end{split}
\ee
where 
\be
\tilde{x}_j^\mu=\frac{x_j^\mu+x_{j-1}^\mu}{2} 
\qquad
\Delta x_j^\mu=x_j^\mu-x_{j-1}^\mu
\ee
and $a_\mu$ and $b$ are functions to be determined by matching
\be\label{dTG}
\begin{split}
2i\partial_T\mathcal{G}&(T,x,y)=\frac{2i}{\delta}\lim_{\delta\to0}\bigg[-\mathcal{G}(T,x,y)\\
&+\int\ud^4 z\sqrt{g(z)}\langle x|e^{-\frac{i\delta}{2}K}|z\rangle\mathcal{G}(T,z,y)\bigg]
\end{split}
\ee
with
\be
2i\partial_T\mathcal{G}(T,x,y)=K_x\mathcal{G}(T,x,y) \;.
\ee
This is the approach that Feynman took in his original paper~\cite{Feynman:1948ur}.
The determinants in the prefactor of~\eqref{matrixAnsatz} give $1+\mathcal{O}(\delta)$, so we could absorb them into the definition of $b$, but keeping them in the prefactor makes it easier to compare with the literature; see e.g. Eq.~(2.25) in~\cite{Grosche:1992qx}. There are similar changes~\cite{Grosche:1992qx} depending on different choices of $\bar{x}_j^\mu$ in $g_{\mu\nu}(\bar{x}_j^\mu)$ and $a_\mu(\bar{x}_j^\mu)$. We will ultimately choose the midpoint $\bar{x}_j^\mu=\tilde{x}_j^\mu$. 

To calculate the $z$ integral in~\eqref{dTG}, we change variable from $z^\mu=x^\mu-\sqrt{\delta}u^\mu$ to $u^\mu$, expand the integrand in powers of $\sqrt{\delta}$ and then perform the resulting Gaussian $u$ integrals. To see what would change if the fields are not evaluated at the midpoint, we perform the initial steps of the calculations with $\tilde{x}=(x+z)/2$ replaced by $x+\lambda(z-x)=x-\lambda\sqrt{\delta} u$, where $\lambda=1/2$ corresponds to the midpoint. We also allow $\lambda$ to be different in the different factors, $g_{\mu\nu}(x-\lambda_0\sqrt{\delta}u)$ and $a_\mu(x-\lambda_1\sqrt{\delta}u)$ [$\delta b(x-\lambda_2\sqrt{\delta}u)\approx\delta b(x)$].
We have to $\mathcal{O}(\delta^2)$ 
\be
e^{-i\delta b/2}=1-\frac{i\delta b}{2} \;,
\ee
\be
\begin{split}
&e^{-i\sqrt{\delta}u^\mu a_\mu(x-\lambda_1\sqrt{\delta} u)}\\
&=1-i\sqrt{\delta}u^\mu a_\mu+\delta u^\mu u^\nu\left[i\lambda_1 a_{\mu,\nu}-\frac{a_\mu a_\nu}{2}\right] \;,
\end{split}
\ee
\be
\frac{\sqrt{g(\tilde{x})}}{[g(x)g(z)]^{1/4}}=1-\frac{\delta}{8}u^\mu u^\nu\Gamma^\rho_{\rho\mu,\nu} \;,
\ee
\be\label{guuExp}
\begin{split}
&e^{-\frac{i}{2}u^\mu u^\nu g_{\mu\nu}(x-\lambda_0\sqrt{\delta} u)}\\
&=\left[1-\lambda_0\sqrt{\delta}u^\mu\partial^x_\mu+\frac{\lambda_0^2\delta}{2}u^\mu u^\nu\partial^x_\mu\partial^x_\nu\right]e^{-\frac{i}{2}u^\mu u^\nu g_{\mu\nu}} 
\end{split}
\ee
and
\be
\begin{split}
\sqrt{g}\mathcal{G}(z)
=\left[1-\sqrt{\delta}u^\mu\partial^x_\mu+\frac{\delta}{2}u^\mu u^\nu\partial^x_\mu\partial^x_\nu\right]\sqrt{g}\mathcal{G}(x) \;.
\end{split}
\ee

There are gamma matrices in $a_\mu$ and $b$, so one might at first worry about the ordering of products of $a$ and $b$ in~\eqref{matrixAnsatz}, but we see from the above expansions that there is only one such product, $u^\mu u^\nu a_\mu(x)a_\nu(x)$, and it is just the square of $a$, so there is no ambiguities in the ordering of $a$ and $b$ in~\eqref{matrixAnsatz}. However, when multiplying all the short-time-step elements $\langle x_j|e^{-i\delta K/2}|x_{j-1}\rangle$ in~\eqref{DiracSteps} we need to proper-time order the products of~\eqref{matrixAnsatz}.  

By performing the $u$ integral before taking the derivatives in~\eqref{guuExp}, we avoid integrals with $u^\alpha u^\beta u^\gamma u^\delta$ and $u^\alpha u^\beta u^\gamma u^\delta u^\epsilon u^\zeta$ in the pre-exponential factor, so we only need the two integrals
\be
\int\frac{\ud^4u}{(2\pi)^2}(1,u^\mu u^\nu)e^{-\frac{i}{2}g_{\alpha\beta}(x)u^\alpha u^\beta}=\frac{1}{\sqrt{g}}(1,-ig^{\mu\nu}) \;.
\ee
Plugging this into~\eqref{dTG} gives
\be\label{dTG2}
2i\partial_T\mathcal{G}=\left(b-\frac{1}{4}g^{\mu\nu}\Gamma^\rho_{\rho\mu,\nu}\right)\mathcal{G}+\frac{g^{\mu\nu}}{\sqrt{g}}K_{\mu\nu}\sqrt{g}\mathcal{G} \;,
\ee
where
\be
K_{\mu\nu}=\mathcal{D}_\mu\mathcal{D}_\nu
+(2\lambda_1-1)ia_{\mu,\nu}+2\lambda_0\overset{\leftarrow}{\partial_\mu}\mathcal{D}_\nu+\lambda_0^2\overset{\leftarrow}{\partial}_\mu\overset{\leftarrow}{\partial}_\nu
\ee
and $\mathcal{D}_\mu=\partial_\mu+ia_\mu$. This already suggests that we should choose $a_\mu$ so that $\mathcal{D}_\mu=D_\mu$, i.e.
\be\label{aSol}
a_\mu=-\frac{1}{4}\omega_{\mu ab}\sigma^{ab}+A_\mu \;,
\ee
and we will see that this does indeed work. From the second term we see that $\lambda_1=1/2$ is a natural choice, so we evaluate $a_\mu$ at the midpoint. By commuting $\sqrt{g}$ past the derivatives using~\eqref{gder} we find
\be
\begin{split}
\frac{g^{\mu\nu}}{\sqrt{g}}K_{\mu\nu}\sqrt{g}=g^{\mu\nu}\Big\{&[D_\mu+\Gamma^\rho_{\rho\mu}][D_\nu+\Gamma^\sigma_{\sigma\nu}]
\\
+&2\lambda_0[\overset{\leftarrow}{\partial_\mu}-\Gamma^\rho_{\rho\mu}][D_\nu+\Gamma^\sigma_{\sigma\nu}]\\
+&\lambda_0^2[\overset{\leftarrow}{\partial}_\mu-\Gamma^\rho_{\rho\mu}][\overset{\leftarrow}{\partial}_\nu-\Gamma^\sigma_{\sigma\nu}]\Big\}
\end{split}
\ee
and after simplifying the $\overset{\leftarrow}{\partial_\mu}\mathcal{D}_\nu$ term using 
\be
\partial_\rho g^{\mu\nu}=-g^{\mu\alpha}\Gamma^\nu_{\alpha\rho}-g^{\nu\alpha}\Gamma^\mu_{\alpha\rho}
\ee
we are also lead to choosing $\lambda_0=1/2$. Thus, we evaluate all the fields in the exponent of~\eqref{matrixAnsatz} at the midpoint $\tilde{x}=(x+z)/2$. 

We then find
\be
\frac{g^{\mu\nu}}{\sqrt{g}}K_{\mu\nu}\sqrt{g}=g^{\mu\nu}(D_\mu D_\nu-\Gamma^\rho_{\mu\nu}D_\rho)+B \;,
\ee
where
\be
\begin{split}
B&=g^{\mu\nu}\left[\frac{3}{4}\Gamma^\rho_{\rho\mu,\nu}-\Gamma^\rho_{\rho\mu}\Gamma^\sigma_{\sigma\nu}-\Gamma^\rho_{\mu\nu}\Gamma^\sigma_{\sigma\rho}\right]+\frac{\sqrt{g}}{4}\partial_\mu\partial_\nu\frac{g^{\mu\nu}}{\sqrt{g}} \\
&=\frac{1}{4}(R+g^{\mu\nu}\Gamma^\rho_{\sigma\mu}\Gamma^\sigma_{\rho\nu})\;.
\end{split}
\ee
Comparing this with~\eqref{KbeforeWeyl} shows that the choices we have made do indeed work, provided that we finally choose
\be\label{bSol}
b=\frac{R}{4}+\frac{1}{2}\sigma^{\mu\nu}F_{\mu\nu}+1+\Delta K \;,
\ee
where $\Delta K$ is a DeWitt term~\cite{DeWitt:1957at}
\be\label{DeWitt}
\Delta K=-B=-\frac{1}{4}(R+g^{\mu\nu}\Gamma^\rho_{\mu\sigma}\Gamma^\sigma_{\nu\rho}) \;.
\ee
This is a term that one would not have guessed if one just followed the usual rule of thumb, which suggests that the exponent in a path integral, $\int\mathcal{D}x\, e^{-iS}$, is just the classical action, $S[x]=S_{\rm cl}[x]$. When comparing with the literature, note that $\Delta K$ depends on the regularization scheme~\cite{Grosche:1992qx,Bastianelli:2006rx,Bastianelli:1998jm,Bastianelli:2000nm}. 
However, when we turn to the instanton approximation in the next section we will be able to neglect the DeWitt term to leading order.

\subsection{Final result}

Inserting~\eqref{matrixAnsatz} into~\eqref{DiracSteps} gives
\be\label{GDiracFin}
\begin{split}
\mathcal{G}(x,y)=\frac{1}{[g(x)g(y)]^{1/4}}\prod_{i=1}^{N-1}\ud^4 x_i\prod_{j=1}^N\frac{\sqrt{g(\tilde{x}_i)}}{(2\pi\delta)^2}\mathcal{P}e^{-iS} \;,
\end{split}
\ee
where $\mathcal{P}$ means proper-time ordering,
\be
S=\sum_{j=1}^N\delta\bigg[g_{\mu\nu}(\tilde{x}_j)\frac{\Delta x_j^\mu\Delta x_j^\nu}{2\delta^2}+\frac{\Delta x_j^\mu}{\delta}a_\mu(\tilde{x}_j)+\frac{b(\tilde{x}_j)}{2}\bigg] \;,
\ee
and $a_\mu$ and $b$ are given by~\eqref{aSol} and~\eqref{bSol}.

In the continuum limit we have
\be
S=\int_{-T/2}^{T/2}\ud u\left[\frac{1}{2}g_{\mu\nu}\dot{x}^\mu\dot{x}^\nu+a_\mu\dot{x}^\mu+\frac{b}{2}\right]
\ee
and the full propagator, obtained from~\eqref{GtildeG} and~\eqref{Tint}, is given by
\be\label{Gfinal}
G(x_\LCp,x_\LCm)=(i\slashed{D}_\LCp+1)\int_0^\infty\frac{\ud T}{2}\int_{x_\LCm}^{x_\LCp}\mathcal{D}x\, \mathcal{P}e^{-iS} \;.
\ee

The amplitude for pair production is obtained by applying the LSZ reduction formula to the propagator,
\be\label{LSZ}
M=\lim_{t_\LCpm\to\infty}\int\ud^3{\bf x}_\LCm\ud^3{\bf x}_\LCm\, e^{ipx_\LCp+iqx_\LCm}\bar{u}\gamma^0 G(x_\LCp,x_\LCm)\gamma^0 v \;.
\ee
We have assumed that the particles eventually end up in flat space at asymptotic times and that it is possible to choose coordinates so that $g_{\mu\nu}=\eta_{\mu\nu}$ there, so $u({\bf p})^{-ipx}$ and $v({\bf q})e^{iqx}$ are free/vacuum wave functions for the particle and antiparticle. We can also replace $\slashed{D}_\LCp\to \slashed{\partial}_\LCp$ and $[g(x_\LCp)g(x_\LCm)]^{-1/4}\to1$ for the same reason.

This assumption is less restrictive than in the SWF approach, because even if we consider metrics with $g_{\mu\nu}(\infty,x)\ne\eta_{\mu\nu}$, such as~\eqref{Aex2D}, in the instanton approximation we still have $\lim_{u\pm\infty}g_{\mu\nu}[t(u),x(u)]\to\eta_{\mu\nu}$ as long as $\dot{x}(\pm\infty)\ne0$ and $g_{\mu\nu}(\infty,\pm\infty)=\eta_{\mu\nu}$.

\section{Worldline instantons}

Having derived a convenient worldline representation in the previous section, we are now ready to turn to the instanton approximation.
To leading order in the semiclassical expansion, we can neglect the DeWitt terms $R$ and $g\Gamma\Gamma$ in~\eqref{bSol},
\be\label{Sinstanton1}
\begin{split}
S=\frac{T}{2}+\int_{-T/2}^{T/2}\ud u\bigg[&\frac{1}{2}g_{\mu\nu}\dot{x}^\mu\dot{x}^\nu+A_\mu\dot{x}^\mu\\
+&\frac{1}{4}\left(F_{\mu\nu}e^\mu_a e^\nu_b-\dot{x}^\mu\omega_{\mu ab}\right)\sigma^{ab}\bigg] \;,
\end{split}
\ee
where the first row is $S_{-1}=\mathcal{O}(1/\hbar)$ and the second row $S_0=\mathcal{O}(\hbar^0)$, while the neglected terms are higher orders. All the integrals will be performed with the saddle-point method (except for those integrals that give delta functions in case $g_{\mu\nu}$ and $A_\mu$ do not depend on all coordinates). The saddle point for the path integral is referred to as an instanton. Only $S_{-1}$ contribute to the saddle-point equations, while $S_0$ is simply evaluated on the saddle points.  

We begin with the path integral.
We change variables and notation as $x^\mu(u)\to x^\mu(u)+\delta x^\mu(u)$, where $\delta x^\mu(u)$ is the new integration variables and $x^\mu(u)$ a saddle point, i.e. an instanton, which is a (complex) solution to the geodesic-Lorentz-force equation
\be\label{geoEq}
\ddot{x}^\mu+\Gamma^\mu_{\rho\sigma}\dot{x}^\rho\dot{x}^\sigma=F^\mu_\nu\dot{x}^\nu
\qquad
F^\mu_{ \nu}=g^{\mu\rho}F_{\rho\nu}\;.
\ee

To perform the $T$ integral with the saddle-point method, we change parametrization from propertime $u$ to a normalized propertime
$\tau$, so that
\be\label{Sminus1}
S_{-1}=\frac{T}{2}+\int_0^1\ud\tau\frac{g_{\mu\nu}\dot{x}^\mu\dot{x}^\nu}{2T}+A_\mu\dot{x}^\mu \;.
\ee
The saddle-point equation for $T$ is then the on-shell condition
\be
T^2=\int\ud\tau\,g_{\mu\nu}\dot{x}^\mu\dot{x}^\nu=g_{\mu\nu}\dot{x}^\mu\dot{x}^\nu \;,
\ee
which after changing back from $\tau$ to $u$ becomes
\be
g_{\mu\nu}\dot{x}^\mu\dot{x}^\nu=1 \;.
\ee
The saddle-point equations for ${\bf x}_\LCm$ and ${\bf x}_\LCp$ give the boundary conditions that determine the instanton solution from the asymptotic momenta of the particle ($p_\mu$) and antiparticle ($q_\mu$)
\be
\dot{x}^\mu(-\infty)=-q^\mu \qquad \dot{x}^\mu(\infty)=p^\mu \;.
\ee
These boundary conditions do not determine the instanton uniquely, so there is a great deal of freedom to choose different instantons (and still obtain the same results).

Plugging the instanton into
\be
\varphi=ipx_\LCp+iqx_\LCm-iS_{-1} \;,
\ee
performing partial integration and using the instanton equation~\eqref{geoEq} gives
\be
\varphi=i\int\ud u\,x^\mu\left(\frac{1}{2}g_{\rho\sigma,\mu}\dot{x}^\rho\dot{x}^\sigma+A_{\rho,\mu}\dot{x}^\rho\right) \;.
\ee
Each instanton gives one term, so the amplitude is given by a sum over all instantons with the same asymptotic momenta ($p_\mu$ and $q_\mu$), 
\be
M=\sum_j\text{pre}_j e^{\varphi_j} \;,
\ee
which can lead to interference effects.
The prefactors are obtained by expanding to second order around the saddle points/instantons and performing the resulting Gaussian integrals, to which we now turn.

\subsection{Functional determinant and geodesic deviation}

Expanding the exponent $S_{-1}$ to second order in $\delta x^\mu$ gives a Gaussian path integral
\be\label{GaussInt}
\int\mathcal{D}\delta x\exp\left\{-\frac{i}{2}\int\ud u\,\delta x^\mu\Lambda_{\mu\nu}\delta x^\nu\right\} \;,
\ee
where
\be
\begin{split}
\Lambda_{\mu\nu}=&-\partial_u g_{\mu\nu}\partial_u+g_{\rho\nu,\mu}\dot{x}^\rho\partial_u-\partial_u g_{\rho\mu,\nu}\dot{x}^\rho
+\frac{g_{\rho\sigma,\mu\nu}}{2}\dot{x}^\rho\dot{x}^\sigma \\
&+\partial_\mu A_\nu\partial_u-\partial_u\partial_\nu A_\mu+A_{\rho,\mu\nu}\dot{x}^\rho
\end{split}
\ee
and the $u$ derivatives act on everything to the right. This form shows that $\Lambda_{\mu\nu}$ is symmetric, i.e. for general $a^\mu(u)$ and $b^\mu(u)$ we have
\be
\int\ud u\, a^\mu\Lambda_{\mu\nu}b^\nu=\int\ud u\, b^\mu\Lambda_{\mu\nu}a^\nu \;.
\ee
Expanding the derivatives and using~\eqref{geoEq} for the $-g_{\rho\mu,\nu}\ddot{x}^\rho$ term gives
\be\label{LambdaDDtoUD}
\Lambda_{\mu\nu}=:g_{\mu\alpha}\Lambda^\alpha_\nu \;,
\ee
where
\be\label{LambdaUD}
\begin{split}
\Lambda^\mu_\nu=&-\left(\delta^\mu_\nu\partial_u^2+2\Gamma^\mu_{\rho\nu}\dot{x}^\rho\partial_u+\Gamma^\mu_{\rho\sigma,\nu}\dot{x}^\rho\dot{x}^\sigma\right) \\
&+F^\mu_\nu\partial_u+F^\mu_{\rho,\nu}\dot{x}^\rho \;,
\end{split}
\ee
where $F^\alpha_{\rho,\nu}=[g^{\alpha\sigma}F_{\sigma\rho}]_{,\nu}$.
By comparing~\eqref{LambdaUD} with~\eqref{geoEq}, we see that $\Lambda^\mu_\nu$ is the same operator that gives the Jacobi or geodesic deviation equation, i.e. if $x^\mu$ and $x^\mu+\phi^\mu$ are two solutions to the geodesic equation~\eqref{geoEq}, then to $\mathcal{O}(\phi)$ we have
\be\label{deviationEq}
\begin{split}
&\Lambda^\mu_\nu\phi^\nu=0 \\
&\ddot{\phi}^\mu+2\Gamma^\mu_{\rho\sigma}\dot{x}^\rho\dot{\phi}^\sigma+\Gamma^\mu_{\rho\sigma,\nu}\dot{x}^\rho\dot{x}^\sigma\phi^\nu=
F^\mu_\nu\dot{\phi}^\nu+F^\mu_{\rho,\nu}\dot{x}^\rho\phi^\nu \;.
\end{split}
\ee
This relation to the geodesic deviation equation is also relevant when we expand the probability around the saddle-point values of the momenta.

As an aside, we note that the geodesic and deviation equations can be expressed in terms of covariant derivatives as
\be
\begin{split}
D\dot{x}^\mu&=F^\mu_\nu\dot{x}^\nu \\
D^2\phi^\mu&=R^\mu_{\alpha\beta\gamma}\dot{x}^\alpha\dot{x}^\beta\phi^\gamma
+F^\mu_\nu D\phi^\nu+D_\nu F^\mu_\rho\dot{x}^\rho\phi^\nu\;,
\end{split}
\ee
where
\be\label{covariantDu}
\begin{split}
D_\nu F^\mu_\rho&=F^\mu_{\rho,\nu}+\Gamma^\mu_{\nu\sigma}F^\sigma_\rho-\Gamma^\sigma_{\rho\nu}F^\mu_\sigma \\
Dv^\mu&=\dot{v}^\mu+\Gamma^\mu_{\rho\sigma}\dot{x}^\rho v^\sigma \;.
\end{split}
\ee

The Gaussian path integral~\eqref{GaussInt} is proportional to $1/\sqrt{\text{Det }\Lambda}$, where $\text{Det }\Lambda$ is a functional determinant. With $\Lambda_{\mu\nu}$ factorized as~\eqref{LambdaDDtoUD}, the determinant also factorizes
\be
\text{Det }\Lambda_{\mu\nu}=\text{Det }g_{\mu\nu}\;\text{Det }\Lambda^\mu_\nu
\ee
and then $1/\sqrt{\text{Det }g_{\mu\nu}}$ cancels against $\prod\sqrt{g}$ in~\eqref{GDiracFin}. This is an important point; in other applications of path integrals in curved space-times, the standard approach involves exponentiating $\prod\sqrt{g}$ in~\eqref{GDiracFin} by introducing additional path integrals~\cite{Bastianelli:2006rx}, but we see now that this is not necessary in this instanton approach.   

Multiplying and dividing by the free path integral then gives
\be
\eqref{GaussInt}=\frac{1}{(2\pi T)^2}\sqrt{\frac{\text{Det }(\Lambda_{\rm free})^\mu_\nu}{\text{Det }\Lambda^\mu_\nu}}
\qquad
(\Lambda_{\rm free})^\mu_\nu=-\delta^\mu_\nu\partial^2 \;.
\ee
The functional determinants can be computed with the Gelfand-Yaglom method~\cite{Dunne:2006st,Dunne:2006ur},
\be
\eqref{GaussInt}=\frac{1}{(2\pi T)^2}\sqrt{\frac{\text{det }\delta x_{\rm free (\nu)}^\mu}{\text{det }\delta x_{(\nu)}^\mu}}\Bigg|_{u=T/2} \;,
\ee
where $\phi^\mu=\delta x_{(\nu)}^\mu$ are 4 solutions to the deviation equation~\eqref{deviationEq} with initial conditions
\be\label{deltaxT}
\delta x_{(\nu)}^\mu(-T/2)=0
\qquad
\delta\dot{x}_{(\nu)}^\mu(-T/2)=\frac{\delta_\nu^\mu}{T} \;.
\ee
We have chosen the normalization so that $\text{det }\delta x_{\rm free (\nu)}^\mu(T/2)=1$.

We will take the limit $t_\LCpm\to\infty$ before starting with any numerical computation. In this limit we also have 
\be
T=u_1-u_0\approx\frac{t_\LCm}{q_0}+\frac{t_\LCp}{p_0}\to\infty \;,
\ee
so we have to replace~\eqref{deltaxT} with something that does not involve $T$. An approach for doing that was developed in~\cite{DegliEsposti:2022yqw,DegliEsposti:2023qqu,Torgrimsson:2025ihd} for the QED case. Two important ingredients in that approach are the following two facts: 1) $\phi^\mu=\dot{x}^\mu$ is a solution to the deviation equation (\eqref{deviationEq} with $g_{\mu\nu}=\eta_{\mu\nu}$), and 2) $\nu=\eta_{\mu\nu}\dot{x}^\mu\dot{\phi}^\nu$ is a constant of motion ($\partial_u\nu=0$). In the GR+QED case, $\phi^\mu=\dot{x}^\mu$ is still a solution to~\eqref{deviationEq}, but now the constant of motion is given by
\be\label{nu}
\nu[\phi]=g_{\mu\nu}\dot{x}^\mu(\dot{\phi}^\nu+\Gamma^\nu_{\rho\sigma}\dot{x}^\rho\phi^\sigma) \;.
\ee
However, at asymptotic times we still have
\be
\nu[\phi]=\eta_{\mu\nu}\dot{x}^\mu\dot{\phi}^\nu(-\infty)=\eta_{\mu\nu}\dot{x}^\mu\dot{\phi}^\nu(\infty) \;,
\ee
so the formulas below take the same form as in the QED case.

In this approach, the solutions with initial conditions~\eqref{deltaxT} at $u=-T/2$ are replaced with a basis of Dirichlet and Neumann solutions defined by
\be\label{DNinit}
\begin{aligned}
D_\mu^\nu(\tilde{u}_0)&=\delta_\mu^\nu
&\qquad
\dot{D}_\mu^\nu(\tilde{u}_0)&=0 \\
N_\mu^\nu(\tilde{u}_0)&=0
&\qquad
\dot{N}_\mu^\nu(\tilde{u}_0)&=\delta_\mu^\nu \;,
\end{aligned}
\ee
where $\tilde{u}_0$ is chosen so that $\ddot{x}^\mu\approx0$ for $\text{Re }u<\text{Re }\tilde{u}_0$, and the lower index is part of the name, so that a general solution can be expressed as a superposition as
\be
\phi^\mu(u)=D^\mu_\nu(u)a^\nu+N^\mu_\nu(u)b^\nu \;,
\ee
where $a^\nu$ and $b^\nu$ are some constant coefficients. In particular, the instanton velocity is a superposition of the Dirichlet solutions alone,
\be
\dot{x}^\mu(u)=-D^\mu_\nu(u)q^\nu \;.
\ee
These properties allow one to factor out the divergent part from the determinant as~\cite{DegliEsposti:2022yqw,DegliEsposti:2023qqu,Torgrimsson:2025ihd}
\be\label{detLambdah}
\text{det }\Lambda=\left(\frac{t_\LCm t_\LCp}{Tq_0p_0}\right)^{D-1}\frac{\tilde{h}}{q_0p_0} \;,
\ee
where $D=2,3,4$ is the number of nontrivial spacetime dimensions ($g_{\mu\nu}(t,x)$, $g_{\mu\nu}(t,x,y)$, $g_{\mu\nu}(t,x,y,z)$) and in 4D 
\be\label{hNum}
\begin{split}
\tilde{h}&=-p_0\text{det}\left[p^\mu,\dot{D}_1^\mu,\dot{D}_2^\mu,\dot{D}_3^\mu\right]\\
&=-\text{det}[\dot{\bf D}_1,\dot{\bf D}_2,\dot{\bf D}_3] \;,
\end{split}
\ee
where $\dot{\bf D}_1=(\dot{D}_1^1,\dot{D}_1^2,\dot{D}_1^3)$.
The $D=1$ case $g_{\mu\nu}(t)$ is special and is treated in Appendix~\ref{detLambdagt}.

\subsection{Functional determinant in 2D}

Eq.~\eqref{hNum} is already suitable for numerical evaluation, but in 2D we can rewrite it as follows, which provides some additional insight. We will basically repeat what was done in~\cite{DegliEsposti:2022yqw,DegliEsposti:2023qqu} for the QED case. We can express an arbitrary solution to the deviation equation as
\be\label{chieta}
(\delta t,\delta x)=(\dot{t},\dot{x})\chi+(-\dot{x},\dot{t})\frac{\alpha}{\sqrt{g}}\frac{\eta}{\dot{t}^2+\dot{x}^2}
\ee
and then work with $\chi(r)$ and $\eta(r)$ instead of $\delta t$ and $\delta x$. $\alpha$ is a constant. We will see below that choosing $\alpha=\dot{t}(\tilde{u}_0)=-q_0$ makes the resulting formulas neat and manifestly symmetric.
If we substitute~\eqref{chieta} into the equation $\delta\ddot{t}=\dots$, solve it algebraically for $\ddot{\chi}$ and substitute that into $\delta\ddot{x}=\dots$, then we obtain a second order equation for $\eta$ that does not involve $\chi$. The coefficient $1/[\sqrt{g}(\dot{t}^2+\dot{x}^2)]$ in~\eqref{chieta} has been chosen to remove any $\dot{\eta}$ term. We find
\be\label{etaeq}
\ddot{\eta}=\frac{R}{2}\eta \;,
\ee
where $R$ is the Ricci scalar.

We have proven~\eqref{etaeq} using Mathematica, by starting with a general 2D metric of the form (with $-\ud y^2-\ud z^2$)
\be
\ud s^2=a(t,x)\ud t^2+2c(t,x)\ud t\ud x-b(t,x)\ud x^2 \;,
\ee
finding $\ddot{\chi}$ using Solve[...], making an educated guess for the coefficient of $\eta$ in~\eqref{etaeq} and then confirming it using Simplify[...]. $R$ involves $(a,b,c)$ and their first and second derivatives with respect to $t$ and $x$. The general expression is long, so we will not present it here, but it simplifies considerably for the two classes of metrics in~\eqref{Rinflation2D} and~\eqref{Ricci_GP}. 

From~\eqref{hNum} we find that the functional determinant only involves $\eta$,
\be
\tilde{h}=-p_0(\dot{t}\delta\dot{x}-\dot{x}\delta\dot{t})=-p_0\alpha\dot{\eta}(\tilde{u}_1) \;,
\ee
so we do not need to find $\chi$. Asymptotically, we have
\be
\alpha\eta=(-\dot{x},\dot{t})\cdot(\delta t,\delta x) \;,
\ee
so with $(\delta t,\delta x)$ given by $D_1^\mu$ and $\alpha=-q_0$, the initial conditions for $\eta$ are
\be\label{eta_in_u0}
\eta(\tilde{u}_0)=1 \qquad \dot{\eta}(\tilde{u}_0)=0 
\ee
and the functional determinant becomes
\be
\frac{\tilde{h}}{q_0p_0}=\dot{\eta}(\tilde{u}_1) \;.
\ee

Since we use initial conditions at $u=0$ for $x^\mu(u)$, it is convenient to write $\eta$ as a superposition 
\be
\eta(u)=a\eta_D(u)+b\eta_N(u)
\ee
of two solutions to~\eqref{etaeq} with initial conditions at $u=0$ rather than at $\tilde{u}_0$,
\be
\begin{aligned}
\eta_D(0)&=1 &\qquad \dot{\eta}_D(0)&=0 \\
\eta_N(0)&=0 &\qquad \dot{\eta}_N(0)&=1 \;.
\end{aligned}
\ee
The constants $a$ and $b$ are found by imposing~\eqref{eta_in_u0}. We find
\be
\dot{\eta}(\tilde{u}_1)=\frac{1}{W(\tilde{u}_0)}[\dot{\eta}_D(\tilde{u}_1)\dot{\eta}_N(\tilde{u}_0)-\dot{\eta}_D(\tilde{u}_0)\dot{\eta}_N(\tilde{u}_1)] \;,
\ee
where $W$ is the Wronskian
\be
W(u)=\eta_D(u)\dot{\eta}_N(u)-\dot{\eta}_D(u)\eta_N(u) \;.
\ee
It follows from~\eqref{etaeq} that $\dot{W}=0$, so
\be
W(\tilde{u}_0)=W(0)=1 \;.
\ee
We arrive at a manifestly symmetric formula for the functional determinant,
\be
\frac{\tilde{h}}{q_0p_0}=\dot{\eta}_D(\tilde{u}_1)\dot{\eta}_N(\tilde{u}_0)-\dot{\eta}_D(\tilde{u}_0)\dot{\eta}_N(\tilde{u}_1) \;.
\ee

\subsection{Ordinary integrals}

The ordinary integrals over ${\bf X}=(T,{\bf x}_\LCm,{\bf x}_\LCp)$ only depend on the asymptotic energies of the particles and are therefore exactly the same as in the QED case. The calculation is described in Appendix~B of~\cite{DegliEsposti:2023qqu}. Expanding around the saddle point, $\delta{\bf X}={\bf X}-{\bf X}_s$, gives a Gaussian integral 
\be
\int\ud^{2D-1}\delta{\bf X}e^{-\delta{\bf X}\cdot{\bf H}\cdot\delta{\bf X}}=\frac{\pi^{(2D-1)/2}}{\sqrt{\text{det }{\bf H}}} \;,
\ee
where
\be
\text{det }{\bf H}=\frac{(q_0p_0)^{D+1}}{2^{2D-1}(t_\LCm t_\LCp)^{D-1}T} \;.
\ee
If the field does not depend on $x^l$, then we have a factor of
\be
\sqrt{2\pi T}2\pi\delta(p'_l+p_l) 
\ee
on the amplitude level.
The 1D case is somewhat special, but can be calculated in the same way~\cite{DegliEsposti:2021its}.

\subsection{Spin part}

The spin part of the propagator is given by $i\slashed{D}_\LCp+1$ in~\eqref{Gfinal} and the second row in~\eqref{Sinstanton1}, 
\be\label{Sigmap}
\Sigma=(\slashed{p}+1)\mathcal{P}\exp\left(-\frac{i}{4}\int\ud u\,M_{ab}\sigma^{ab}\right) \;,
\ee
where
\be\label{MFomega}
M_{ab}=F_{\mu\nu}e^\mu_a e^\nu_b-\dot{x}^\mu\omega_{\mu ab} 
\ee
is evaluated on the instanton solution (so both $u$ and $M_{ab}$ are in general complex). 
At first glance, \eqref{Sigmap} does not look symmetric: $\slashed{p}+1$ contains the asymptotic particle momentum, but there is no similar term for the antiparticle. We can prove that it is in fact symmetric as in~\cite{Torgrimsson:2025ihd}, i.e. by defining  
\be
\Sigma(u,u_0)=[\gamma_\mu\dot{x}^\mu(u)+1]\mathcal{P}\exp\left(-\frac{i}{4}\int_{u_0}^u M_{ab}\sigma^{ab}\right) 
\ee
and studying $\partial_u\Sigma$. 

In the following formulas, $v^\mu$ is either the instanton velocity $v^\mu=\dot{x}^\mu$ or a spin vector $v^\mu=\alpha^\mu$.
We have ($D$ is given by~\eqref{covariantDu})
\be\label{duslv}
\partial_u\slashed{v}=\partial_u(v^\mu e_{\mu a}\gamma^a)=\gamma_\mu Dv^\mu+\dot{x}^\mu\omega_{\mu ab}v^a\gamma^b \;,
\ee
which we obtained using~\eqref{connection2}.
From the derivative of the exponential we have a factor of $M_{ab}\sigma^{ab}$ which we commute to the left using
\be\label{slvM}
\left[\slashed{v},-\frac{i}{4}M_{ab}\sigma^{ab}\right]
=-\gamma_\mu F^\mu_\nu v^\nu-\dot{x}^\mu\omega_{\mu ab}v^a\gamma^b \;.
\ee
Since $x^\mu(u)$ is a solution to~\eqref{geoEq}, the right-hand sides of~\eqref{duslv} and~\eqref{slvM} cancel and we find
\be\label{diffEqForSpin}
\partial_u\Sigma(u,u_0)=-\frac{i}{4}M_{ab}(u)\sigma^{ab}\Sigma(u,u_0) \;, 
\ee
which allows us to express $\Sigma(u,u_0)$ as
\be
\Sigma(u,u_0)=\mathcal{P}\exp\left(-\frac{i}{4}\int_{u_0}^u M_{ab}\sigma^{ab}\right)\Sigma(u_0,u_0) \;.
\ee
But 
\be
\Sigma(u_0,u_0)=\gamma_\mu\dot{x}^\mu(u_0)+1\to-\slashed{q}+1
\ee
as $\text{Re }u_0\to-\infty$, so~\eqref{Sigmap} can also be expressed as
\be
\Sigma=\mathcal{P}\exp\left(-\frac{i}{4}\int\ud u\,M_{ab}\sigma^{ab}\right)(-\slashed{q}+1) \;,
\ee
which shows that we do indeed have a particle-antiparticle symmetry, even though it is not immediately obvious.
The fact that the cancellations in this calculation worked out also provides a check of the spin term in~\eqref{Sinstanton1}. 

Using $(1+\slashed{p})^2=2(1+\slashed{p})$ we can write
\be
\Sigma=\frac{1}{2}(1+\slashed{p})\mathcal{P}e^{-\frac{i}{4}\int M\sigma}(-\slashed{q}+1) 
\ee
and plugging this into~\eqref{LSZ} gives
\be\label{ubPev}
\bar{u}\gamma^0\Sigma\gamma^0v=-2p_0q_0\bar{u}\mathcal{P}e^{-\frac{i}{4}\int M\sigma}v \;.
\ee

In some cases one can simplify the path-ordered exponential analytically. Even if this is not possible, \eqref{diffEqForSpin} is trivial to solve numerically. There is therefore no reason to try to avoid path-ordered exponentials by introducing Grassmann path integrals for these instanton calculations.

We can find simple explicit expressions for any metric of the form~\eqref{inflation2D}, because the integrand in~\eqref{Sigmap} can be written as a total derivative (cf. the QED case in~\cite{DegliEsposti:2022yqw})
\be
\dot{x}^1\omega_{1\hat{0}\hat{1}}=-s\partial_u\ln[\dot{t}+s e_1^1\dot{x}^1] 
\qquad
s=\pm1
\;,
\ee
so
\be\label{spinExp2D}
\frac{i}{4}2\int\ud u\,\dot{x}\omega_{1\hat{0}\hat{1}}\sigma^{\hat{0}\hat{1}}=\gamma^{\hat{0}}\gamma^{\hat{1}}\ln L \;,
\ee
where
\be
\begin{split}
\ln L&=s\frac{1}{2}\ln\left(\dot{t}+s e^1_1\dot{x}\right)\Big|^\infty_{-\infty}\\
&=\ln\left[i\epsilon\sqrt{\frac{p^0+p^1}{q^0+q^1}}\right]
=\ln\left[i\epsilon\sqrt{\frac{q^0-q^1}{p^0-p^1}}\right] \;,
\end{split}
\ee
where $\epsilon=\pm1$. This is exactly the same as in QED, so, with spinors chosen as 
\be\label{uv}
u_s({\bf p})=\frac{(1+\slashed{p})R_s}{\sqrt{2p_0(p_0+p_1)}}
\quad
v_s({\bf p})=\frac{(1-\slashed{p})R_s}{\sqrt{2p_0(p_0+p_1)}} \;,
\ee
where
\be\label{Reqs}
\gamma^{\hat{0}}\gamma^{\hat{1}}R_s=R_s
\qquad
i\gamma^{\hat{2}}\gamma^{\hat{3}} R_s=sR_s 
\qquad
R_r^\dagger R_s=\delta_{rs} \;,
\ee
we find a simple background-independent result,
\be\label{u0Sigma0v2D}
\bar{u}_r\gamma^0\Sigma\gamma^0v_s=-2\sqrt{p_0q_0}i\epsilon\delta_{rs} \;.
\ee

Motivated by this 2D result~\eqref{u0Sigma0v2D}, we define a conveniently normalized spin factor as
\be\label{Srs}
\begin{split}
S_{rs}&=\frac{1}{2\sqrt{p_0q_0}}\bar{u}_r\gamma^0\Sigma\gamma^0v_s \\
&=-\sqrt{p_0q_0}\bar{u}_r\mathcal{P}e^{-\frac{i}{4}\int M\sigma}v_s\;,
\end{split}
\ee
and now return to a general 4D background.
We can describe the spin in terms of a 4-vector $\alpha^\mu(u)$, which is orthogonal to the instanton velocity and normalized as
\be
g_{\mu\nu}\dot{x}^\mu\alpha^\nu=0
\qquad
g_{\mu\nu}\alpha^\mu\alpha^\nu=-1 \;,
\ee
and evolves according to the generalization of the Bargmann-Michel-Telegdi (BMT) equation~\cite{Bargmann:1959gz} to curved spacetime,
\be\label{BMT}
\dot{\alpha}^\mu+\Gamma^\mu_{\rho\sigma}\dot{x}^\rho\alpha^\sigma=F^\mu_\nu\alpha^\nu \;.
\ee
Since $[\dot{x}_\mu\gamma^\mu,\gamma^5\slashed{\alpha}]=0$ and $(\gamma^5\slashed{\alpha})^2=1$, we can choose the spinors to be eigenstates with eigenvalues $s=\pm1$ as
\be\label{alphaBasis}
\gamma^5\slashed{\alpha}(u_0)v_s({\bf q})=sv_s({\bf q})
\qquad
\gamma^5\slashed{\alpha}(u_1)u_s({\bf p})=su_s({\bf p}) \;.
\ee
We can choose an arbitrary antiparticle spin vector $\alpha_{(0)}^\mu=\alpha^\mu(u_0)$, restricted only by $\alpha_{(0)}^2=-1$ and $q\alpha_{(0)}=0$, but then the particle spin vector $\alpha^\mu(u_1)$ in~\eqref{alphaBasis} is determined by transporting $\alpha_{(0)}^\mu$ along the instanton trajectory according to~\eqref{BMT}; or vice versa. To show that this is indeed a convenient choice, we first use~\eqref{duslv} and~\eqref{slvM} with $v^\mu=\alpha^\mu$ to show that $\Sigma$ transforms $\gamma^5\slashed{\alpha}(u_0)$ to $\gamma^5\slashed{\alpha}(u_1)$ as
\be
\Sigma(u_1,u_0)\gamma^5\slashed{\alpha}(u_0)=\gamma^5\slashed{\alpha}(u_1)\Sigma(u_1,u_0) \;.
\ee
Using this in~\eqref{ubPev} gives
\be\label{diagonal1}
\begin{split}
&\bar{u}_r\mathcal{P}e^{-\frac{i}{4}\int M\sigma}v_s=
s\bar{u}_r\mathcal{P}e^{-\frac{i}{4}\int M\sigma}\gamma^5\slashed{\alpha}v_s=\\
&s\bar{u}_r\gamma^5\slashed{\alpha}\mathcal{P}e^{-\frac{i}{4}\int M\sigma}v_s
=rs\bar{u}_r\mathcal{P}e^{-\frac{i}{4}\int M\sigma}v_s\propto\delta_{rs}f(s) \;.
\end{split}
\ee
From this we can conclude that the spin factor is diagonal in a basis chosen as~\eqref{alphaBasis},
\be
S_{rs}(\alpha)=\begin{pmatrix} a(\alpha)&0\\0&b(\alpha) \end{pmatrix} \;,
\ee
for some $a$ and $b$. This holds for any choice of $\alpha^\mu(u_0)$. Denote two arbitrary spin vectors as $\alpha_\mu$ and $\alpha'_\mu$. The spinors $u'$ and $v'$ defined using $\alpha'_\mu$ must be related to the spinors $u$ and $v$ defined with $\alpha^\mu$ via a unitary transformation, i.e. there must exist a unitary matrix $U(u)$ such that
\be
v'_{s'}=U_{s's}(u_0)v_s
\qquad
u'_{s'}=U_{s's}(u_1)u_s \;.
\ee
This transforms the spin factor as
\be\label{UcSUT}
S_{rs}(\alpha')=[U^*(u_1)S(\alpha)U^T(u_0)] \;.
\ee
But $S_{rs}(\alpha')$ must still be diagonal. Demanding that the off-diagonal elements of~\eqref{UcSUT} vanish gives conditions that allow us to determine some of the elements of $U(u_0)$ in terms of $U(u_1)$. This is enough to simplify the diagonal elements, and we find
\be
S_{rs}=S_0\begin{pmatrix} e^{i\phi_1}&0\\0& e^{i\phi_2}\end{pmatrix} \;,
\ee
where $S_0$ is independent of $\alpha$ and the phases $\phi_i\in\mathbb{R}$ are undetermined. We can absorb these phases into the undetermined phases of the particle or antiparticle states, e.g. $e^{i\phi_s}v_s\to v_s$. Thus, up to an arbitrary choice of phases for the states, we find for any $\alpha$ 
\be
S_{rs}=S_0\delta_{rs} \;.
\ee
Remembering the extra minus sign that comes from normal ordering the antiparticle mode operators, $N(dd^\dagger)=-d^\dagger d$, $v_\LCp$ actually corresponds to spin down and $v_\LCm$ to spin up, so $S_{rs}=S_0\delta_{rs}$ means equal probability to produce $|\uparrow\downarrow\rangle$ and $|\downarrow\uparrow\rangle$ and zero probability to produce $|\uparrow\uparrow\rangle$ or $|\downarrow\downarrow\rangle$. This holds for any choice of spin basis, as long as the spin axes for the particle and antiparticle are related according to the BMT equation~\eqref{BMT}.

\subsection{Final formulas}

Thus, given a set of instantons, which in general have to be found numerically, the pair-production spectrum is obtained with the same formulas as in the QED case, namely
\be\label{P112131}
\begin{split}
3+1:&\quad P=\int\frac{\ud^3{\bf p}\ud^3{\bf q}}{(2\pi)^3}|\hat{M}|^2 \\
2+1:&\quad P=V_z\int\frac{\ud^3{\bf p}\ud q_1\ud q_2}{(2\pi)^3}|\hat{M}|^2\Big|_{q_3=-p_3} \\
1+1:&\quad P=V_yV_z\int\frac{\ud^3{\bf p}\ud q_1}{(2\pi)^3}|\hat{M}|^2\Big|_{q_{2,3}=-p_{2,3}} \\
0+1:&\quad P=V_xV_yV_z\int\frac{\ud^3{\bf p}}{(2\pi)^3}|\hat{M}|^2\Big|_{{\bf q}=-{\bf p}} \;,
\end{split}
\ee
where ${\bf p}$ and ${\bf q}$ are the particle and antiparticle momenta, $V_i$ is a volume factor for $x^i$, the (normalized) amplitude is a sum with one term for each instanton,
\be\label{Msum}
\hat{M}=\sum_j\hat{M}_j=\sum_j\frac{S_je^{\varphi_j}}{\tilde{h}_j^{1/2}} \;,
\ee
where $\tilde{h}$ comes from the functional determinant~\eqref{Sigmap}, except for the special $0+1$D case where $\tilde{h}\to1$ in~\eqref{Msum}, as shown in Appendix~\ref{detLambdagt}, 
and the spin factor is given by~\eqref{Srs}.

\subsection{Grid vs. quadratic instanton approaches}

We have two approaches to study the momentum spectrum: the grid approach, where we make a grid in the $({\bf p},{\bf q})$ space and calculate instantons at each grid point, $x^\mu(u;{\bf p},{\bf q})$; and the quadratic approach, where we calculate instantons only for the saddle-point values of $({\bf p},{\bf q})$, $x^\mu(u;{\bf p}_s,{\bf q}_s)$, and expand the exponential parts of the amplitude terms in~\eqref{Msum} to second order in ${\bm\Pi}-{\bm\Pi}_s$, where ${\bm\Pi}=({\bf p},{\bf q})$, 
\be\label{quadMexp}
\begin{split}
&M_j({\bm\Pi})\approx\frac{S_j}{\tilde{h}_j^{1/2}}({\bm\Pi}_s) \\
&\times e^{\varphi_j({\bm\Pi}_s)+\nabla\varphi_j\cdot({\bm\Pi}-{\bm\Pi}_s)
+\frac{1}{2}({\bm\Pi}-{\bm\Pi}_s)\cdot\nabla\nabla\varphi_j\cdot({\bm\Pi}-{\bm\Pi}_s)} \;.
\end{split}
\ee
We choose ${\bm\Pi}_s$ such that $\text{Re }\nabla\varphi({\bm\Pi}_s)=0$, so ${\bm\Pi}_s$ is not actually a saddle point, unless there is only one instanton, in which case the imaginary part of $\varphi$ drops out in $|M|^2$. The actual saddle points for $\varphi_j$ are in general complex~\cite{DegliEsposti:2024upq}.

Since we assume that the particles end up in Minkowski space, the formulas for $\nabla\varphi$ and $\nabla\nabla\varphi$ in~\eqref{quadMexp} are identical to the QED case~\cite{DegliEsposti:2022yqw,DegliEsposti:2023qqu}, so the following equations have been simply copied from~\cite{Torgrimsson:2025ihd}. 
For the first-order derivatives we only need the instanton $x^\mu$ but not any deviation solution $\delta x^\mu$,  
\be\label{firstOrderDer}
\begin{split}
\frac{\partial\varphi}{\partial q^k}&=-i\left(x^k-\frac{q^k}{q_0}t\right)(\tilde{u}_0)\\
\frac{\partial\varphi}{\partial p^k}&=-i\left(x^k-\frac{p^k}{p_0}t\right)(\tilde{u}_1) \;.
\end{split}
\ee
For the second-order derivatives we also need solutions to the geodesic-deviation equation with boundary conditions
\be\label{deltaBC}
\begin{aligned}
\delta\dot{x}_{[l]}^k(-\infty)&=\delta_{kl}
&\qquad
\delta\dot{x}_{[l]}^k(\infty)&=0 \\
\delta\dot{x}_{\{l\}}^k(-\infty)&=0
&\qquad
\delta\dot{x}_{\{l\}}^k(\infty)&=-\delta_{kl} \;.
\end{aligned}
\ee
We have
\be
\begin{split}
\frac{\partial^2\varphi}{\partial q^k\partial q^l}&=i\left(\delta x_{[l]}^k-\frac{q^k}{q_0}\delta t_{[l]}+\frac{q_0^2\delta_{kl}-q^kq^l}{q_0^3}t\right)(\tilde{u}_0)\\
&=i\left(\delta x_{[k]}^l-\frac{q^l}{q_0}\delta t_{[k]}+\frac{q_0^2\delta_{kl}-q^kq^l}{q_0^3}t\right)(\tilde{u}_0)
\end{split}
\ee
\be
\begin{split}
\frac{\partial^2\varphi}{\partial p^k\partial p^l}&=i\left(\delta x_{\{l\}}^k-\frac{p^k}{p_0}\delta t_{\{l\}}+\frac{p_0^2\delta_{kl}-p^kp^l}{p_0^3}t\right)(\tilde{u}_1)\\
&=i\left(\delta x_{\{k\}}^l-\frac{p^l}{p_0}\delta t_{\{k\}}+\frac{p_0^2\delta_{kl}-p^kp^l}{p_0^3}t\right)(\tilde{u}_1)
\end{split}
\ee
\be
\begin{split}
\frac{\partial^2\varphi}{\partial q^k\partial p^l}&=i\left(\delta x_{\{l\}}^k-\frac{q^k}{q_0}\delta t_{\{l\}}\right)(\tilde{u}_0)\\
&=i\left(\delta x_{[k]}^l-\frac{p^l}{p_0}\delta t_{[k]}\right)(\tilde{u}_1) \;.
\end{split}
\ee

The results obtained with this quadratic-instanton approximation are compared with the exact SWF results in Figs.~\ref{fig:gauss_2D_03}, \ref{fig:gauss_1D_03}, \ref{fig:minus_gauss_2D_03}, \ref{fig:minus_gauss_1D_03}, \ref{fig:large_d} and~\ref{fig:holonomy2D}. The relative errors are consistent with the general expectation that they scale as $\mathcal{O}(\omega)$, corresponding to $\mathcal{O}(10\%)$ for the values of $\omega$ considered. For more complicated backgrounds, it may be advantageous to construct a grid-based instanton approximation before resorting to SWF. However, for the 2D examples considered above, the full 2D spectra (e.g. Figs.~\ref{fig:gauss_2D_03} and~\ref{fig:minus_gauss_2D_03}) can be computed in less than a minute on an A100 GPU, which is significantly faster than the time required to set up the grid-instanton calculation. In fact, even the quadratic-instanton approximation is slower than the SWF computation in these cases, since each new background requires determining suitable initial guesses for $\dot{x}^\mu(0)$ to ensure convergence of the Newton-Raphson method, as well as choosing appropriate contours to navigate branch points in the complex $u$ plane. By contrast, in the SWF approach one only needs to ensure sufficiently dense and wide grids in $x$ and $k$. The advantage of the instanton approximation becomes apparent in higher-dimensional settings, or when parameters are taken away from $\mathcal{O}(1)$, since in those cases the wave functions can vary rapidly, making the SWF approach more challenging. 

Even when the grid-instanton approach requires substantially more manual effort, it would still be desirable to demonstrate that it provides better agreement with the SWF results than the quadratic-instanton approximation, as was done in~\cite{Torgrimsson:2025pao,Torgrimsson:2025ihd} for the QED case. This turned out to be more challenging for the GR examples considered here. Here there are always at least two instantons for each $({\bf p},{\bf q})$ because gravity does not distinguish particles and antiparticles. The two classes of instantons correspond to the choices $\dot{x}(0)=+i$ or $\dot{x}(0)=-i$. For the simplest metrics, this leads to a decomposition of the amplitude, $M(P,\Delta)=M_\LCm(P,\Delta)+M_\LCp(P,\Delta)$, where $M_\LCpm(P,\Delta)$ are peaked at $P=\pm P_s$, $\Delta=0$. If these peaks were sufficiently well separated, so that $|M|^2\approx|M_\LCm|^2+|M_\LCp|^2$, it would be straightforward to compute the grid-instanton approximation. However, in most of the examples considered here there is significant overlap, so the cross term cannot be neglected. As a consequence, one must evaluate $M_\LCm(P,\Delta)$ not only for $P<0$ but also for $P>0$. It has turned out to be nontrivial to continue across $P=0$. Typically, when moving in $(P,\Delta)$ space, one can use the same $u$ contour, or deform it smoothly when branch points approach the contour. However, at $P=0$ it may be necessary to change the contour discontinuously, without guidance from the previous contour. This issue is left for future work.

\section{Conclusions}

We have shown that the SWF approach, originally introduced in~\cite{Torgrimsson:2025pao,Torgrimsson:2025ihd} for pair production in electromagnetic background fields, can also be applied to pair production in curved spacetimes with nontrivial dependence on two or more coordinates. We have also completed the development of an open-worldline-instanton method by calculating the pre-exponential part of the probability, complementing the calculation of the exponential contribution presented in~\cite{Semren:2025dix}. We have compared the results obtained with these two methods and found good agreement.

Although more involved than 1D backgrounds, the 2D examples studied in this work remain relatively simple and were chosen primarily to test the new formulas and their implementation. To facilitate a direct comparison between the two approaches, we have focused on parameter regimes in which both methods are comparatively straightforward to apply. For the SWF approach, this meant restricting attention to parameter values of $\mathcal{O}(1)$, while for the instanton approach we selected cases with only a small number of relevant instantons. In this regime, the exact SWF method can in fact be faster than the approximate instanton approach.

The computational cost of the instanton method itself is typically low. The equations to be solved are ordinary differential equations for $x^\mu(u)$ and $\delta x^\mu(u)$, and the Newton--Raphson procedure used to determine $x^\mu(0)$ (and $\dot{x}^\mu(0)$ in higher dimensions) generally converges within only a few iterations when supplied with suitable starting points. However, identifying such starting points often requires a non-negligible amount of manual work. One possible strategy is to introduce a deformation $g_{\mu\nu}(t,x)\to g_{\mu\nu}(t,\beta x)$ and begin with $\beta=0$, where the relevant instantons are much easier to identify. The solutions obtained in this limit can then serve as initial guesses for $\beta=1$, possibly through a sequence of intermediate values of $\beta$ if required.

A further challenge is the choice of contour for the proper time $u$. In general, the contour cannot simply coincide with the real axis, but must instead pass around branch points in the complex plane in an appropriate manner. As long as the contour remains on the same side of these branch points, it may be continuously deformed without affecting the resulting probability. We have found it convenient to represent the contour as a sequence of straight-line segments,
\be
u(r)=u(r_j)+(r-r_j)e^{i\phi_j}
\qquad
r_j<r<r_{j+1} \;,
\ee
with constant angles $\phi_j$. For the QED examples studied so far, a three-segment contour has usually been sufficient, with $\phi=0$ for the antiparticle asymptotic region ($\text{Re }u\to-\infty$), $\phi=-\pi/2$ in the creation region, and $\phi=0$ again for the particle asymptotic region ($\text{Re }u\to+\infty$). For the gravitational examples considered here and in~\cite{Semren:2025dix}, however, it has often been necessary to choose $\phi\neq-\pi/2$ or even introduce additional contour segments in the creation region. Moreover, changes in the background parameters can shift the locations of branch points, requiring corresponding modifications of the contour. 

By contrast, applying the SWF method to different metrics requires comparatively little manual work. One just needs to adjust the density and width of the grids in $x$ and $k$ to achieve convergence and the desired numerical precision. For the simple 2D examples considered here, the SWF code runs very efficiently on modern hardware; on an NVIDIA A100 GPU (which we have used via Google Colab), a typical calculation for 2D plots such as Figs.~\ref{fig:gauss_2D_03} and~\ref{fig:minus_gauss_2D_03}, requires less than a minute. For such backgrounds, the SWF approach can therefore outperform the instanton method in terms of overall turnaround time.

The situation changes, however, when moving away from $\mathcal{O}(1)$ parameter values or extending the problem to 3D or 4D. Although additional integration constants ($x^\mu(0)$ and $\dot{x}^\mu(0)$) must be determined in 3D and 4D, the instanton equations still depend on only a single independent variable (proper time $u$). In contrast, large parameter values can lead to rapidly oscillating wave functions, making the SWF approach increasingly demanding. Furthermore, the transition from 2D to 3D or 4D increases the dimensionality of the grids in both ${\bf x}$ and ${\bf k}$, leading not only to longer runtimes but also to substantial memory requirements. In~\cite{Torgrimsson:2025ihd}, this issue was addressed in the QED case by splitting the $(p_1,q_1)$ grid into smaller batches. Unlike the Dirac--Heisenberg--Wigner method~\cite{Bialynicki-Birula:1991jwl}, each point on the $(p_1,q_1)$ grid corresponds to an independent computation in the SWF approach, naturally lending itself to parallelization over multiple GPUs in future implementations.

Several directions for future work therefore suggest themselves. On the instanton side, it would be desirable to automate the identification of initial conditions and contour choices, thereby reducing the amount of manual input required. On the SWF side, further development of memory-efficient implementations, including parallelization strategies and techniques inspired by recent advances in machine learning, could significantly extend the range of tractable multidimensional problems.

\acknowledgements

G.T. is supported by Olle Engkvists Stiftelse.

\appendix

\section{Two methods for metrics with nontrivial holonomy}\label{twoMethods}

\subsection{Method 1}

When the holonomy is trivial, i.e. when we can choose coordinates so that $g_{\mu\nu}\to\eta_{\mu\nu}$ everywhere in the exterior region, then we can compute the inner products by first separating out a delta-function part as in~\eqref{deltaSeparated}.
If instead $\hat{y}^a\ne\check{y}^a$ then it is more nontrivial to extract the delta-function parts of $\langle U_{\rm in}|U_{\rm out}\rangle$ and $\langle V_{\rm in}|V_{\rm out}\rangle$. As an integral over a finite interval of $x$ cannot give a delta function, there cannot be any delta function in $\langle U_{\rm in}|U_{\rm scat}\rangle$ or $\langle V_{\rm in}|V_{\rm scat}\rangle$, because $\psi_{\rm scat}(t_{\rm in},x)$ is approximately zero for $|x|$ larger than some finite $L$, and the delta functions in $\langle U_{\rm in}|U_{\rm back}\rangle$ and $\langle V_{\rm in}|V_{\rm back}\rangle$ must come from the asymptotic regions $|x|\gg1$. 

Consider for simplicity the example in~\eqref{inflation2D} and let $h=e^{A(t,x)}$, then (suppressing the spin indices)
\be
\begin{split}
\langle U_{\rm in}(k_1)|U_{\rm back}(p_1)\rangle&=u_k^\dagger u_p e^{i(k_0-p_0)t_\LCm}J(k_1,p_1) \\
\langle U_{\rm in}(k_1)|V_{\rm back}(q_1)\rangle&=u_k^\dagger v_q e^{i(k_0+q_0)t_\LCm}J(k_1,-q_1) \\
\langle V_{\rm in}(-k_1)|U_{\rm back}(p_1)\rangle&=v_{-k_1}^\dagger u_p e^{i(-k_0-p_0)t_\LCm}J(k_1,p_1) \\
\langle V_{\rm in}(-k_1)|V_{\rm back}(q_1)\rangle&=v_{-k_1}^\dagger v_q e^{i(-k_0+q_0)t_\LCm}J(k_1,-q_1)\;,
\end{split}
\ee
where
\be
J(k_1,p_1)=\int\ud\check{x}\,\exp[i(k_1-p_1)\check{x}+i\mathcal{P}(\check{x},p_1)] \;,
\ee
\be\label{checkxh}
\check{x}(x)=\int_0^x\ud\tilde{x}\,h(-\infty,\tilde{x})
\ee
and
\be\label{Phh}
\mathcal{P}=p_1[\check{x}-\hat{x}(\check{x})] \;.
\ee
This $J$ is of the same form as that in~\cite{Torgrimsson:2025pao}, but with a different $\mathcal{P}$, so we can reuse the following result from~\cite{Torgrimsson:2025pao},
\be\label{Jdeltapv}
J(k,p)=2\pi\delta(\Delta)\frac{1}{2}\left[e^{i\mathcal{P}(-\infty)}+e^{i\mathcal{P}(\infty)}\right]-\text{p.v.}\frac{\tilde{J}}{\Delta} \;,
\ee
where $\Delta=k_1-p_1$, $\text{p.v.}$ means principal value and
\be
\tilde{J}=\int_{-\infty}^\infty\ud\check{x}\,\mathcal{P}'(\check{x})\exp\left\{i\Delta\check{x}+i\mathcal{P}(\check{x})\right\} \;.
\ee
For~\eqref{Phh} we have
\be
\mathcal{P}(\pm\infty,p_1)=p_1\int_0^{\pm\infty}\!\ud x[h(-\infty,x)-h(\infty,x)]
\ee
and
\be
\mathcal{P}'(\check{x})=p_1[1-\hat{x}'(\check{x})]=p_1\left[1-\frac{h(\infty,x[\check{x}])}{h(-\infty,x[\check{x}])}\right] \;,
\ee
where $x(\check{x})$, the inverse of~\eqref{checkxh}, can be obtained by solving the differential equation
\be
x'(\check{x})=1/h[-\infty,x(\check{x})]
\qquad x(0)=0 \;.
\ee
Thus, the inner products split into
\be\label{JUU}
\begin{split}
&\langle U_{\rm in}(w,k)|U_{\rm out}(s,p)\rangle \\
&=2\pi\delta(k_1-p_1)\delta_{ws}\frac{1}{2}\left[e^{i\mathcal{P}(-\infty,p_1)}+e^{i\mathcal{P}(\infty,p_1)}\right]\\
&-\text{p.v.}\frac{\tilde{J}(k_1,p_1)}{k_1-p_1}u_k^\dagger u_p e^{i(k_0-p_0)t_\LCm}\\
&+\langle U_{\rm in}(w,k)|U_{\rm scat}(s,p)\rangle  \;,
\end{split}
\ee
\be\label{JVU}
\begin{split}
\langle V_{\rm in}(w,-k_1)|U_{\rm out}(s,p_1)\rangle 
&=-\frac{\tilde{J}(k_1,p_1)}{k_1-p_1}v_{-k}^\dagger u_p e^{i(-k_0-p_0)t_\LCm}\\
&+\langle V_{\rm in}(w,-k_1)|U_{\rm scat}(s,p_1)\rangle \;,
\end{split}
\ee
\be\label{JUV}
\begin{split}
\langle U_{\rm in}(w,k)|V_{\rm out}(r,q)\rangle 
&=-\frac{\tilde{J}(k_1,-q_1)}{k_1+q_1}u_k^\dagger v_q e^{i(k_0+q_0)t_\LCm}\\
&+\langle U_{\rm in}(w,k)|V_{\rm scat}(r,q)\rangle \;,
\end{split}
\ee
and
\be\label{JVV}
\begin{split}
&\langle V_{\rm in}(w,-k_1)|V_{\rm out}(r,q_1)\rangle \\
&=2\pi\delta(k_1+q_1)\delta_{wr}\frac{1}{2}\left[e^{i\mathcal{P}(-\infty,-q_1)}+e^{i\mathcal{P}(\infty,-q_1)}\right]\\
&-\text{p.v.}\frac{\tilde{J}(k_1,-q_1)}{k_1+q_1}v_{-k}^\dagger v_q e^{i(-k_0+q_0)t_\LCm}\\
&+\langle V_{\rm in}(w,-k_1)|V_{\rm scat}(r,q_1)\rangle \;.
\end{split}
\ee
For $\langle U_{\rm in}|V_{\rm out}\rangle$ and $\langle V_{\rm in}|U_{\rm out}\rangle$ there is no need for $\text{p.v.}$ because
\be
\lim_{k_1\to p_1}\frac{v_{-k}^\dagger u_p}{k_1-p_1}=\frac{1}{2p_0^2}
\qquad
\lim_{k_1\to-q_1}\frac{u_k^\dagger v_q}{k_1+q_1}=\frac{1}{2q_0^2} \;.
\ee
After computing $\psi_{\rm scat}$, Fourier transforming and computing~\eqref{JUU}, \eqref{JVU}, \eqref{JUV} and~\eqref{JVV}, 
we can now perform the $k_1$ integral in
\be
\begin{split}
&\int\frac{\ud k_1}{2\pi}[\langle U_{\rm in}(k)|U_{\rm out}(p)\rangle]^\dagger \langle U_{\rm in}(k)|V_{\rm out}(q)\rangle \\
=-&\int\frac{\ud k_1}{2\pi}[\langle V_{\rm in}(k)|U_{\rm out}(p)\rangle]^\dagger \langle V_{\rm in}(k)|V_{\rm out}(q)\rangle \;.
\end{split}
\ee
Computing both lines gives a useful check of the numerical implementation. The principal-value integrals can be computed as
\be
\text{p.v.}\int_{-\infty}^\infty\!\frac{\ud\Delta}{\Delta}f(\Delta)=\int_0^\infty\frac{\ud\Delta}{\Delta}[f(\Delta)-f(-\Delta)] \;.
\ee

\subsection{Method 2}

For the second method, we begin by defining
\be\label{curly}
\begin{split}
\{U_{\rm in}|\psi\}(\check{x})&=\int\frac{\ud k_1}{2\pi}e^{-ik_1\check{x}}u_k^\dagger\int\ud\check{y}e^{ik_1\check{y}}\psi(\check{y})\\
\{V_{\rm in}|\psi\}(\check{x})&=\int\frac{\ud k_1}{2\pi}e^{-ik_1\check{x}}v_{-k}^\dagger\int\ud\check{y}e^{ik_1\check{y}}\psi(\check{y}) \;,
\end{split}
\ee
which is a Fourier transform, projection with $u_k^\dagger$ or $v_{-k}^\dagger$ and then an inverse Fourier transform. This allows us to express the inner products as
\be
\begin{split}
\langle U_{\rm in}(k)|\psi\rangle&=\int\ud\check{x}\,e^{ik_0 t_\LCm+ik_1\check{x}}\{U_{\rm in}|\psi\}(\check{x})\\
\langle V_{\rm in}(-k)|\psi\rangle&=\int\ud\check{x}\,e^{-ik_0 t_\LCm+ik_1\check{x}}\{V_{\rm in}|\psi\}(\check{x})
\end{split}
\ee
and transform the $k_1$ integral into a $\check{x}$ integral,
\be\label{KtoXintegral}
\begin{split}
&\int\frac{\ud k_1}{2\pi}\langle U_{\rm out}|U_{\rm in}\rangle\langle U_{\rm in}|V_{\rm out}\rangle\\
=&\int\ud\check{x}\,\{U_{\rm in}|U_{\rm out}\}^\dagger(\check{x})\{U_{\rm in}|V_{\rm out}\}(\check{x}) \;.
\end{split}
\ee
The $\check{x}$ integral in~\eqref{KtoXintegral} has no singularities and can be performed simply with e.g. the Simpson's rule.
For $\psi=U_{\rm scat}$ and $\psi=V_{\rm scat}$ we can compute the Fourier and inverse-Fourier integrals in~\eqref{curly} with FFT. 

The background terms can be computed as follows.
We split $\{U_{\rm in}|U_{\rm back}\}$ into two terms by writing $u_k^\dagger u_p=\delta_{ws}+(u_k^\dagger u_p-\delta_{ws})$. 
We have again written out $\delta_{ws}$ as a suggestion for how to generalize to higher dimensions, but for 2D fields and with 2D spinors we only have one spin state and $\delta_{ws}\to1$. The first term has trivial integrals and the second can be rewritten using partial integration, which gives 
\be\label{curlyUinUback}
\begin{split}
\{U_{\rm in}|&U_{\rm back}\}(\check{x})=\delta_{ws}e^{-ip_0t_\LCm-ip_1\hat{x}(\check{x})}\\
+&\int\frac{\ud k_1}{2\pi}e^{-ik_1\check{x}}p_1\frac{u_k^\dagger u_p-\delta_{ws}}{k_1-p_1}\\
\times&\int\ud\check{y}e^{ik_1\check{y}}(\hat{x}'(\check{y})-1)e^{-ip_0t_\LCm-ip_1\hat{x}(\check{x})} \;.
\end{split}
\ee
The integrals in~\eqref{curlyUinUback} can be performed by FFT, multiplication by $p_1(u_k^\dagger u_p-\delta_{ws})/(k_1-p_1)$ and then inverse FFT. Since $u_{-q}^\dagger v_q=0$ we have
\be\label{curlyUinVback}
\begin{split}
\{U_{\rm in}|V_{\rm back}\}(\check{x})=&\int\frac{\ud k_1}{2\pi}e^{-ik_1\check{x}}(-q_1)\frac{u_k^\dagger v_q}{k_1+q_1}\\
\times&\int\ud\check{y}e^{ik_1\check{y}}(\hat{x}'(\check{y})-1)e^{iq_0t_\LCm+iq_1\hat{x}(\check{x})} \;.
\end{split}
\ee
Similar expressions hold for $\{V_{\rm in}|U_{\rm back}\}$ and $\{V_{\rm in}|V_{\rm back}\}$.

\section{$\text{det }\Lambda$ for $g_{\mu\nu}(t)$}\label{detLambdagt}

In this appendix we calculate $\text{det }\Lambda$ for the special 1D case where $A_\mu=0$ and
\be
\ud s^2=\ud t^2-e^{2A(t)}\ud x^2 \;.
\ee
This 1D metric has a couple of peculiar features that are not present in higher-dimensional metrics, $g_{\mu\nu}(t,x,...)$. 

We can integrate the geodesic equation to find the velocity $\dot{x}^\mu$ in terms of $t$ as in~\cite{Semren:2025dix},
\be\label{txSol10}
\dot{t}=\sqrt{1+p^2 e^{-2A(t)}}
\qquad
\dot{x}^1=\dot{x}=p e^{-2A(t)} \;.
\ee
The instanton goes around a branch point so that $\text{Re }\dot{t}<0$ for $u\to u_0$ and $\text{Re }\dot{t}>0$ for $u\to u_1$, so $\sqrt{...}$ in~\eqref{txSol10} denotes the double-valued square root. The reason for the change in sign of $\text{Re }\dot{t}$ is that both particles are final-state particles, so both $\text{Re }t_\LCm\to+\infty$ and $\text{Re }t_\LCp\to+\infty$. 

Since the instantons and geodesic deviations for $g_{\mu\nu}(t)$ have a different asymptotic behavior compared to $g_{\mu\nu}(t,x,,...)$, we cannot use~\eqref{detLambdah} for this case. Instead we go back to the deviation equation~\eqref{deviationEq} and solve it with the following ansatz,
\be
\phi^\mu=a\dot{x}^\mu+b\delta_1^\mu \;,
\ee
where $\delta_1^\mu$ is a Kronecker delta. $\dot{x}^\mu$ is always a solution, and $\delta_1^\mu$ is a solution since $g_{\mu\nu}(t)$ does not depend on $x$, which implies that the differential equations for $\phi$ will only involve terms with derivatives on $a$ and $b$. The constant of motion~\eqref{nu} becomes
\be
\nu=\dot{a}-p\dot{b} \;,
\ee
so
\be
a(u)=\nu(u-u_0)+p b(u) \;.
\ee
From the $\mu=1$ component of~\eqref{deviationEq} we then find
\be
\dot{b}=\beta\frac{e^{-2A(t)}}{\dot{t}^2} \;,
\ee
where $\beta$ is an integration constant. Changing integration variable from $u$ to $t$ and imposing $b(t_\LCm)=0$ at $t_\LCm=t(u_0)$ gives
\be
b=\beta\int_{t_\LCm}^t\ud\tilde{t}\frac{e^{-2A}}{(1+p^2 e^{-2A})^{3/2}} \;.
\ee
The initial conditions
\be
\dot{\phi}_0^\mu(u_0)=\frac{\delta^\mu_0}{T}
\qquad
\dot{\phi}_1^\mu(u_0)=\frac{\delta^\mu_1}{T}
\ee
determine the remaining two integration constants:
\be
\phi_0:
\qquad
(\nu_0,\beta_0)=\frac{\dot{t}(u_0)}{T}(1,-p)
\ee
and
\be
\phi_1:
\qquad
(\nu_1,\beta_1)=\frac{1}{T}(-p,p^2+e^{2A_\infty}) \;,
\ee
where $A_\infty=A(t_\LCm)=A(t_\LCp)$.

In the limit $t_\LCpm\to\infty$ we have
\be
b(u_1)\approx\beta(t_\LCm+t_\LCp)\frac{e^{-2A_\infty}}{(1+p^2e^{-2A_\infty})^{3/2}}
\ee
and hence
\be
\text{det }\Lambda=\phi_1^1\phi_2^2-\phi_1^2\phi_2^1=\frac{t_\LCm+t_\LCp}{T\dot{t}(\infty)}=1 \;.
\ee
This is the same result as in~\cite{DegliEsposti:2021its} for the QED case for $E(t)$. Thus, the functional determinant is trivial for $g_{\mu\nu}(t)$.

\end{document}